\documentclass[12pt, draftclsnofoot, onecolumn]{IEEEtran}
\normalsize
\usepackage{pifont,bm,multicol,amsfonts,amsmath,color,amssymb,graphicx, epsfig,cite,psfrag,subfigure,algorithm,enumerate,stfloats,algorithm,algorithmic,epstopdf,balance}

\begin{document}
\title{Uplink-aided High Mobility Downlink Channel Estimation over Massive MIMO-OTFS System} 	
\author{Yushan Liu, Shun Zhang, {\emph{Member, IEEE,}}  Feifei Gao, {\emph{Fellow, IEEE,}} \\
Jianpeng Ma, {\emph{Member, IEEE,}} Xianbin Wang, {\emph{Fellow, IEEE}}

\thanks{Y. Liu, S. Zhang and J. Ma are with the State Key Laboratory of Integrated Services Networks, Xidian University, Xi’an 710071, P. R. China (Email: $\text{ysliu}\_$97@stu.xidian.edu.cn, zhangshunsdu@xidian.edu.cn, jpmaxdu@gmail.com).}

    \thanks{F. Gao is with Department of Automation, Tsinghua University, State Key Lab of Intelligent Technologies and Systems, Tsinghua University, State Key for Information Science and Technology (TNList) Beijing 100084, P. R. China (Email: feifeigao@ieee.org).}

    \thanks{X. Wang is with Department of Electrical and Computer Engineering, Western University, London, Ontario, Canada (Email: xianbin.wang@uwo.ca).}

}
\maketitle
\vspace{-10mm}

  \begin{abstract}
Although it is often used in the orthogonal frequency division multiplexing (OFDM) systems, application of massive multiple-input multiple-output (MIMO) over the orthogonal time frequency space (OTFS) modulation could suffer from enormous training overhead in high mobility scenarios.
In this paper, we propose one uplink-aided high mobility downlink channel estimation scheme for the massive MIMO-OTFS networks.
Specifically, we firstly formulate the time domain massive MIMO-OTFS signal model along the uplink and adopt the expectation maximization based variational Bayesian (EM-VB) framework to recover the uplink channel parameters including the angle, the delay, the Doppler frequency, and the channel gain for each physical scattering path.
Correspondingly, with the help of the fast Bayesian inference, one low complex approach is constructed to overcome the bottleneck of the EM-VB.
Then, we fully exploit the angle, delay and Doppler reciprocity between the uplink and the downlink and reconstruct the angles, the delays, and the Doppler frequencies
 for the downlink massive channels at the base station.
Furthermore, we examine the downlink massive MIMO channel estimation over
the delay-Doppler-angle domain. The channel dispersion of the OTFS over the delay-Doppler domain is carefully analyzed
and is utilized to associate one given path with one specific delay-Doppler grid if different paths of any user have distinguished
delay-Doppler signatures.
Moreover, when all the paths  of any user could be perfectly separated over the angle domain, we design
the effective path scheduling algorithm to map different users' data into the orthogonal delay-Doppler-angle domain resource
and achieve the parallel and low complex downlink  3D channel estimation.
For the general case, we adopt the least square estimator
with reduced dimension to capture the downlink delay-Doppler-angle channels.
Various numerical examples are presented to confirm the validity and robustness of the proposed scheme.

\end{abstract}

\maketitle
\thispagestyle{empty}
\vspace{-1mm}


\section{Introduction}

Massive multiple-input multiple-output (MIMO) has become one of the most important enabling technologies for the 5-th generation (5G) and beyond wireless networks, due to its tremendously improved spectral and energy efficiencies \cite{massive_MIMO,efficiency2,Massive_in_5G_1,Massive_in_5G_2,Matthaiou_MIMO,JIN_dft}.
To fully exploit the advantages of massive MIMO, a set of designed strategies have been proposed in reducing its implementation cost and complexity \cite{JSDM,JSDM_Opportunistic,ZJY_efficiency, JSDM_mm,Xie2018CCM,reconstr_estimation_FDD,channel_estimation,Matthaiou_massive,gao_E}.
However, in the high mobility scene, most of these methods do not work effectively due to the rapid channel variation.

To overcome this challenge, there are many works about the time-varying massive MIMO channels.
Qin \emph{et al.} proposed one effective time-varying massive MIMO channel estimation scheme for the orthogonal frequency division multiplexing (OFDM) system, where the complex exponential basis expansion model (CE-BEM) was utilized for the representation of the time-varying channels \cite{OFDM_timevarying}.
In \cite{Zhao_timevarying}, Zhao \emph{et al.} designed a channel tracking method for massive MIMO systems under both time-varying and spatial-varying circumstances,
where the effective dimension of the uplink (UL)/ the downlink (DL) channel was reduced by the spatial-temporal basis expansion model.
In \cite{Gao_Highmobility}, Guo \emph{et al.} applied angle domain Doppler compensation for high-mobility wideband massive MIMO UL communications,
where the Doppler spread of the equivalent UL channel after angle domain Doppler compensation was theoretically analyzed.
In \cite{Ma_J_SBL_Time_varing}, Ma \emph{et al.} developed an expectation maximization (EM) based sparse Bayesian learning (SBL) framework to learn the spatial and temporal parameters for the time-varying massive MIMO channel model and applied a Kalman filter (KF) with reduced dimension for the channel tracking.
Furthermore, in \cite{MYLi_SBL_Time_varing}, Li \emph{et al.} expanded the work in \cite{Ma_J_SBL_Time_varing}, where they considered the randomness of direction of arrivals (DOAs), focused on the imperfection of the reconstructed parameters and
proposed an optimal Bayesian Kalman filtering (OBKF) method along DL.
However, The aforementioned methods only consider the block-fading channel, i.e., the channel only changes from block to block.
This assumption is reasonable in low velocity scenario, but may be not applicable in the high speed scenarios.

\begin{figure}[!t]
	\centering
	\includegraphics[width=4.3in]{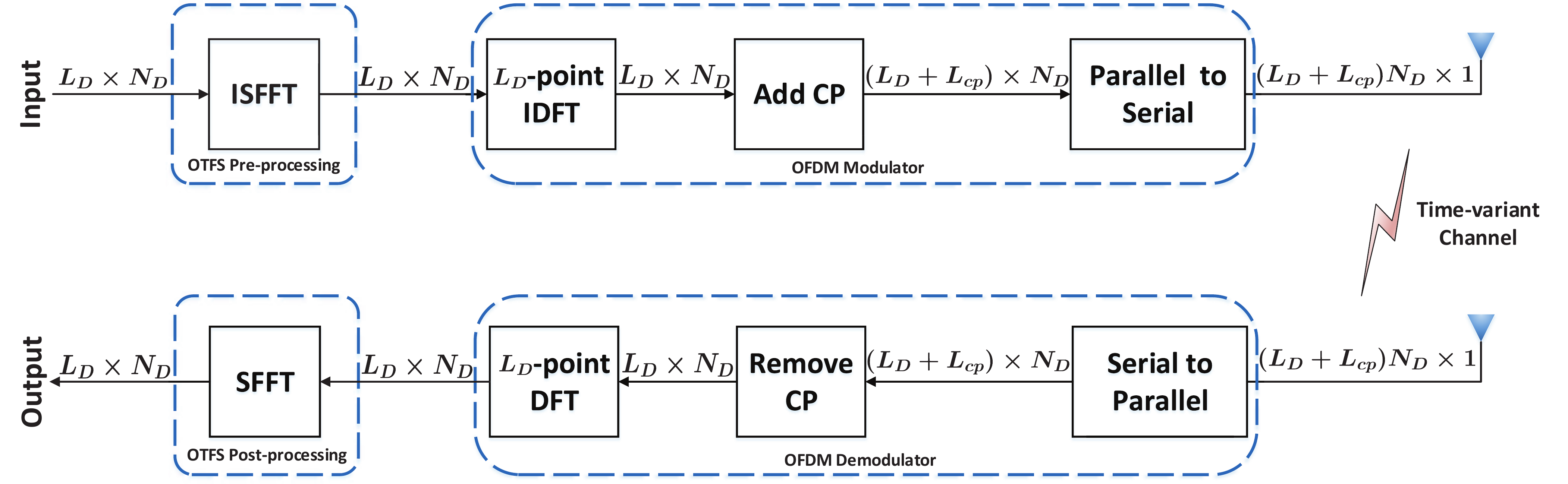}
	\caption{The modulation and demodulation for OTFS.}
	\label{fig:OTFS_modulation_demodulation}
\end{figure}

Meanwhile, the massive MIMO channels may experience the frequency-selective fading, and OFDM is usually adopted.
However, for the high-mobility scenarios, OFDM may possess the significant inter-carrier interference (ICI) due to the Doppler spread of the time-variant channels, which then  severely degrades the system performance.
To deal with {this problem},
Hadani \emph{et al.} designed a novel two-dimensional modulation technique called orthogonal time frequency space (OTFS) modulation \cite{OTFSori}.
Fig. \ref{fig:OTFS_modulation_demodulation} shows the OTFS architecture \cite{OTFSSISO}.
{As the newly proposed  modulation/demodulation technique,}
OTFS can be {achieved through adding some function blocks to the OFDM scheme, i.e.,}  adding a pre-processing block before a traditional OFDM modulator and {a corresponding post-processing block after a traditional OFDM demodulator}.
With the help of the pre-processing and post-processing blocks, {the} time-variant channels are converted into the time-independent channels in the delay-Doppler domain.
Therefore, the information bearing data can be multiplexed {over the} roughly constant channels
{in the delay-Doppler domain}.
At the same time, the {transmitted} data in OTFS systems can take full advantage of the diversity {for} the frequency-time channels.
In this way, OTFS can {improve the} system performance over OFDM in the high-mobility scenarios \cite{OTFSSISO}.
Hence, OTFS has attracted many researchers' attentions in recent years.

Murali \emph{et al.} investigated OTFS modulation for communication over high mobility  channels \cite{OTFSori2}.
They utilized the markov chain monte carlo (MCMC) sampling techniques to construct a low complex detection method and a PN pilot sequence based channel estimation scheme in the delay-Doppler domain.
In \cite{OTFSesti1}, Raviteja \emph{et al.} proposed embedded pilot-aided channel estimation schemes for OTFS.
Raviteja \emph{et al.} derived the explicit input-output relation and developed one novel low complex message passing algorithm for joint interference cancellation and symbol detection \cite{IC}.
In \cite{OTFSesti2}, Shen \emph{et al.} proposed a 3D sparse signal model for the DL massive MIMO channel estimation, which was formulated as the sparse signal recovery problem.

In this paper, we apply OTFS for the massive MIMO network.
Within the traditional massive MIMO-OFDM, the digital precoders can be placed at different sub-carriers
to capture
the channel diversity in the frequency domain.
Nevertheless, the digital precoders for the massive MIMO-OTFS may happen in the delay-Doppler domain. 
In this scenario, the base station (BS) should achieve the channel state information (CSI) in the
delay-Doppler domain. Moreover, the users also need the corresponding CSI to implement decoding in the delay-Doppler domain.
Hence, we will focus on the channel estimation along DL over the massive MIMO-OTFS.
Firstly, we introduce the high mobility massive MIMO channel model and the massive MIMO-OTFS scheme. Then, we develop the time domain massive MIMO-OTFS signal model for the UL, where the angle off-grid is considered.
The expectation maximization based variational Bayesian (EM-VB) framework
is adopted to recover the UL channel parameters including the angles, the delays, the Doppler frequencies, and the channel gains for all scattering paths. To avoid the huge complexity caused by the large matrix inversion,
we design the low complex EM-VB to fully exploit the fast Bayesian inference.
Thanks to the angle, delay, and Doppler reciprocity between the UL and the DL even in the frequency division duplex (FDD) mode,
we can reconstruct the DL parameters including the angles, the delays, and the Doppler frequencies at BS.
Furthermore, we examine the DL channel estimation over the delay-Doppler-angle domain.
The case, where different scattering paths {of any user} possess distinguished
delay-Doppler signatures, is examined to illustrate the channel dispersion property of the OTFS over
the delay-Doppler domain. {Under the scenario where all paths of any user are perfectly separated over the angle domain,
a effective path scheduling algorithm is proposed to parallel implement the DL 3D channel recovering for the multiple users.}
For the {general  case,} we utilize the achieved delay-Doppler-angle signatures to construct
one least square (LS) estimator with reduced dimension to capture the DL delay-Doppler-angle channels.

The rest of this paper is organized as follows. Section II introduces high mobility massive MIMO channels and describes the massive MIMO-OTFS scheme. In Section III,
we extract UL channel parameters with the EM-VB method and construct the low complex EM-VB. In Section IV,
we recover the DL channel parameters at BS and analyze the DL massive MIMO channel estimation over the delay-Doppler-angle domain. The simulation results are presented in Section V, and conclusions are drawn in Section VI.

Notations: We use lowercase (uppercase)  boldface to denote vector (matrix).
$(\cdot)^T$, $(\cdot)^*$, and $(\cdot)^H $ represent the transpose, the
complex conjugate and the Hermitian transpose, respectively. $\mathbf{I}_N$
represents a $N\times N$ identity matrix. {$\mathbf{1}_N$ represents a $N\times 1$ all-one vector.} $\delta(\cdot)$ is the Dirac delta function.  $\mathbb E \{\cdot\}$ is the expectation operator. We use $\text{tr}\{\cdot\}$, $\det\{\cdot\}$ and $\text{rank}\{\cdot\}$ to denote the trace, the determinant, and the  rank of a matrix, respectively. $[\mathbf{X}]_{ij}$ is the $(i,j)$-th entry of $\mathbf{X}$. $[\mathbf X]_{:,\mathcal Q}$ (or $[\mathbf X]_{\mathcal Q,:}$) is the submatrix of $\mathbf{X}$  {and contains} the columns (or rows) with  {the} index set $\mathcal Q$. $\mathbf x_{\mathcal Q}$ is the sub-vector of $\mathbf{x}$ formed by the entries with the index set $\mathcal Q$. {$|\mathcal Q|$ denotes that all elements of the set $\mathcal Q$ take their absolute value. $\|\mathcal Q\|$ represents the number of elements in the set $\mathcal Q$. $\|\mathbf x\|$ denotes the modulus of the vector $\mathbf x$.} $\mathbf{n} \sim \mathcal{CN}(0,\mathbf{I}_{N})$ means that $\mathbf{n}$ is complex circularly-symmetric Gaussian distributed with zero mean and covariance $\mathbf{I}_{N}$. $\lfloor x \rfloor$ denotes the smallest integer no less than $x$, while $\lceil x \rceil$ represents the largest integer no more than $x$. $\backslash$ is the set subtraction operation. $\Re(x)$ is the real component of $x$ and $\Im(x)$ is the image component of $x$.
$\text{diag}(\mathbf X)$ is a column vector formed by the diagonal elements of $\mathbf X$.

\section{System Model}
\subsection{High Mobility Massive MIMO Channel Model}


In this work, we consider {a single-cell massive MIMO system} in the high-mobility scenarios.
The BS serves $K$ users who are randomly distributed in the cell. The BS is equipped with a uniform linear array (ULA), which contains $M$ antenna elements and $M\gg K$, and each user is equipped with single antenna. The wireless signal can reach the user side along the line of sight paths or can be reflected by multiple scatterers,
which means that the channel links between the BS and the users subject to the frequency-selective fading.
Due to the users' mobility, the channels vary with respect to the time, i.e., the channel links experience the time-selective fading. Along DL, we assume there are $P$ scattering paths,
and each scatter path corresponds
to one direction of departure (DOD), one Doppler frequency shift, and one time delay.
Denote $\theta_{k,p}(r)$ as a DOD for the $p$-th path of user $k$ at time $r$,
and the corresponding antenna array spatial steering vector can be defined~as:
\begin{equation}\label{steer_vec}
\mathbf{a}(\theta_{k,p}(r))=[1, e^{\jmath 2\pi \frac{d\sin(\theta_{k,p}(r))}{\lambda}},\ldots,e^{\jmath 2\pi (M-1) \frac{d\sin(\theta_{k,p}(r))}{\lambda}}]^T,
\end{equation}
where $d$ is antenna spacing of the BS, and $\lambda$  is the carrier wavelength for DL.
Hence,
from the geometric channel model,
the time-varying DL channels at time $r$ between BS and the user $k$ can be denoted by
\begin{align}
\mathbf h_{k,l}(r)=\sum_{p=1}^P {h}_{k,p}e^{\jmath2\pi\nu_{k,p}rT_s} \delta(lT_s-\tau_{k,p})
\mathbf a(\theta_{k,p}(r)),\label{eq:h_kln}
\end{align}
where $h_{k,p}$, $\tau_{k,p}$ and $\nu_{k,p}$ represent the channel gain, delay and Doppler shift for
the $p$-th path of the user $k$, respectively, {the index $l$ denotes the index along the delay domain, $r$ represents
the time index,} $\delta(\cdot)$ denotes the Dirac delta function, and $T_s$ is the system sampling period. Furthermore, we suppose that $\tau_{k,p}=r_1T_s$, where $r_1$ is one integer number.
However, the $\theta_{k,p}(r)$ keeps relatively constant within  a  quite long time interval.
For example, let us suppose the user moves at the speed of $200$ km/h, and the distance between BS and user is $500$ m. Within $15$ ms, the user can move only $0.83$ m at most.
Then the change of DoD seen by the BS is less than $0.1^{\circ}$, which is quite small.
Thus, we can omit the time index $r$ of the angle.
Obviously, $\mathbf h_{k,l}(r)$ can be determined by the $P$ parameter sets $\{\tau_{k,p}, \nu_{k,p}, \theta_{k,p},
h_{k,p}\}_{p=1}^P$, and each parameter set corresponds to one propagation path.

\subsection{Massive MIMO-OTFS Scheme}

\subsubsection{OTFS Modulation}
At the $m$-th antenna, a data sequence of length $L_DN_D$ is rearranged into a two-dimensional data block
$\mathbf{X}_m\in\mathcal{C}^{L_D\times N_D}$, where $L_D$ and $N_D$ are the dimensions of the delay domain and the Doppler domain, respectively. And it is called a two-dimensional OTFS block in the delay-Doppler domain.

First, we apply the  inverse symmetric finite Fourier transform (ISFFT) for the  pre-processing block and obtain the data block $\tilde{\mathbf{X}}_m$ in the time-frequency domain as
$\tilde{\mathbf{X}}_m=\mathbf{F}_{L_D}\mathbf{X}_m\mathbf{F}_{N_D}^H$,
where $\mathbf{F}_{L_D}\in\mathcal{C}^{L_D\times L_D}$ and $\mathbf{F}_{N_D}\in\mathcal{C}^{N_D\times N_D}$ are
normalized discrete Fourier transform (DFT) matrices, and their entries can be denoted as $[\mathbf F_{N_D}]_{p,q}=\frac{1}{\sqrt{N_D}} e^{-\jmath\frac{2\pi pq}{N_D}}$.

Then, 
taking the $L_D$-point inverse DFT (IDFT) on each column of $\tilde{\mathbf{X}}_m$, we can obtain the transmitting signal block $\mathbf{S}_m\!\in\!\mathcal{C}^{L_D\times N_D}$ as
$\mathbf{S}_m
\!=\!\mathbf{F}_{L_D}^H\tilde{\mathbf{X}}_m$,
where $\mathbf{S}_m\!=\!\left[\mathbf{s}_{m,0},\cdots,\mathbf{s}_{m,N_D-1}\right]$ and $\mathbf{s}_{m,j}$
{represents the $L_D\times 1$ vector}.
Each column of $\mathbf{S}_m$ can be regarded as an OFDM symbol.
Thus, with respect to one specific delay-Doppler block $\mathbf{X}_m$,
there are $N_D$ OFDM symbols $\{\mathbf{s}_{m,j}\}^{N_D-1}_{j=0}$.
Moreover, the transmitted signal block $\mathbf{S}_m$ can be rewritten as $\mathbf{S}_m=\mathbf{X}_m\mathbf{F}^H_{N_D}$.

In order to avoid the inter-symbol interference between blocks, the OFDM modulator usually adds cyclic prefix (CP) for each OFDM symbol $\mathbf{s}_{m,j}$. Therefore, we can obtain the one-dimensional transmitting signal $\mathbf{s}_m \in\mathcal{C}^{(L_D+L_{cp})N_D\times 1}$ over the time domain as $\mathbf{s}_m=\text{vec}\{\mathbf{A}_{cp}\mathbf{S}_m\}$,
where $\mathbf{A}_{cp}=\big[[\mathbf{I}_{L_D}]_{L_D-L_{cp}:L_D-1,:}^T,\mathbf{I}_{L_D}^T\big]^T\in\mathcal{C}^{(L_D+L_{cp})\times L_D}$ is the CP addition matrix,
and $L_{cp}$ is the length of CP.
Then, $\mathbf{s}_m$ occupies bandwidth $L_D\bigtriangleup f$ with duration $N_DT$,
where $\bigtriangleup f$ and $T=(L_D+L_{cp})T_s$ are the subcarrier spacing and the OFDM symbol period, respectively.
\subsubsection{OTFS Demodulation}
After $\mathbf{s}_m$ passing through the time-varying DL channels, user $k$ can receive the $(L_D+L_{cp})N_D\times 1$ signal vector $\mathbf{z}_k$.
Firstly, we rearrange $\mathbf{z}_k$ as a two-dimensional matrix $\mathbf{Z}_k\in\mathcal{C}^{(L_D+L_{cp})\times N_D}$,~i.e., $\mathbf{Z}_k=\text{unvec}\{\mathbf{z}_k\}$,
where each column vector of $\mathbf{Z}_k$ can be regarded as one received OFDM symbol with CP.
Then, multiplying $\mathbf{Z}_k$ with the CP removal matrix $\mathbf{A}_{rcp}=\left[\textbf{0}_{L_D\times L_{cp}},\mathbf{I}_{L_D}\right]$, we can obtain the OFDM received symbols $\mathbf{A}_{rcp}\mathbf{Z}_k$ without CPs.
With the $L_D$-point DFT on the OFDM received symbols, the received two-dimensional block $\tilde{\mathbf{Y}}_k$ in the time-frequency domain can be written~as $
\tilde{\mathbf{Y}}_k=\mathbf{F}_{L_D}\mathbf{A}_{rcp}\mathbf{Z}_k$.

Finally, with {the SFFT} operation in post-processing block, $\tilde{\mathbf{Y}}_k$ is transformed to the two-dimensional data block $\mathbf{Y}_k\in\mathcal{C}^{L_D\times N_D}$ in the delay-Doppler domain as $ \mathbf{Y}_k=\mathbf{F}_{L_D}^H\tilde{\mathbf{Y}}_k\mathbf{F}_{N_D}$.
Then the received two-dimensional block $\tilde{\mathbf{Y}}_k$ can be rewritten as $ \mathbf{Y}_k=\mathbf{A}_{rcp}\mathbf{Z}_k\mathbf{F}_{N_D}$.

\subsubsection{Simple Representation of the Received Signal}
According to \cite{OTFSesti2}, the $(i,j+N_D/2)$-th entry of $\mathbf{Y}_k$, i.e., $y_{k,i,j+N_D/2}$, can be denoted as
\begin{align}
&[\mathbf Y_k]_{i,j+N_D/2}\approx
\sum_{m=0}^{M-1}\sum_{i'=0}^{L_D-1} \!\!\sum_{j'=-N_D/2}^{N_D/2-1}\!\!\! x_{i',j'+N_D/2,m}\tilde{h}_{k,(i -i')_{L_D},\langle j-j'\rangle,m}e^{\jmath2\pi\frac{i(j-j')}{N_D(L_D+L_{cp})}}\!+\!w_{k,i,j+N_D/2},
\label{eq:fiveA Y^DD}
\end{align}
where $(i\!-\!i')_{L_D}$ is the remainder after division of $i\!-\!i'$ by $L_D$, $\langle j\!-\!j'\rangle$ denotes $(j\!-\!j'\!+\!N_D/2)_{N_D}\!-\!N_D/2$, $x_{i',j'\!+\!N_D/2,m}$ is the $(i',j'\!+\!N_D/2)$-th element of $\mathbf X_m$, and
$\tilde{h}_{k,i,j,m}$ is the equivalent channel over the delay-Doppler-space domain,
 $i\!=\!0,1,\ldots,L_D\!-\!1$, $j\!=\!-N_D/2,\ldots,0,\ldots,N_D/2\!-\!1$.
Here, it is assumed that $w_{k,i,j\!+\!N_D/2}$ is complex Gaussian distributed with zero mean and variance
$\sigma^2$, and is independent from element to element.
Moveover, $\tilde{h}_{k,i,j,m}$ can be derived from (\ref{eq:h_kln}) as
\begin{align}
\tilde{h}_{k,i,j,m}=&\frac{1}{N_D}\sum_{n=1}^{N_D}\left[\mathbf{h}_{k,i}\left((n-1)(L_D+L_{cp})+1\right)\right]_m e^{-\jmath2\pi(n-1)\frac{j}{N_D}}\notag\\
=& \frac{1}{N_D}\sum_{p=1}^Ph_{k,p}e^{\jmath2\pi\nu_{k,p}T_s}
\frac{\sin(\pi (\nu_{k,p} N_DT-j))}{\sin(\pi\frac{(\nu_{k,p}N_DT-j)}{N_D})} e^{\jmath\pi\frac{(\nu_{k,p}N_DT-j)(N_D-1)}{N_D}}\delta(iT_s-\tau_{k,p}) e^{\jmath2\pi m\frac{d\sin\theta_{k,p}}{\lambda}}.\label{eq:fiveB H^DDS}
\end{align}
Similar to \cite{MYLi_SBL_Time_varing, Ma_J_SBL_Time_varing,DFT}, we can utilize the spatial DFT operation
to dig the channel sparsity caused by the massive antennas. Explicitly, taking the normalized DFT along the antenna index $m$,
we can derive the delay-Doppler-angle domain channel $\bar{h}_{k,i,j,q}$ $(q\!=\!-\!\frac{M}{2},\ldots,0,\ldots,\frac{M}{2}\!-\!1)$ as
\begin{align}
\bar{h}_{k,i,j,q}\triangleq&\frac{1}{\sqrt M}\sum_{m=0}^{M-1}\tilde{h}_{k,i,j,m}e^{-\jmath2\pi \frac{q m}{M}},\notag\\
=&\frac{1}{N_D\sqrt M} \sum_{p=1}^Ph_{k,p} e^{\jmath2\pi\nu_{k,p}T_s} \frac{\sin(\pi (\nu_{k,p}N_DT-j))}{\sin(\pi\frac{(\nu_{k,p}N_DT-j)}{N_D})}e^{\jmath\pi\frac{(\nu_{k,p}N_DT-j)(N_D-1)}{N_D}} \notag\\
&\times  \delta(iT_s-\tau_{k,p}) \frac{\sin(\pi(M\frac{d\sin\theta_{k,p}}{\lambda}-q))}{\sin(\pi\frac{M\frac{d\sin\theta_{k,p}}{\lambda}-q}{M})} e^{\jmath\pi\frac{(M\!\frac{d\sin\theta_{k,p}}{\lambda}-q)(M-1)}{M}}.
\label{eq:fiveB H^DDA}
\end{align}

\begin{figure}[!t]
	\centering
	\includegraphics[width=3.4in]{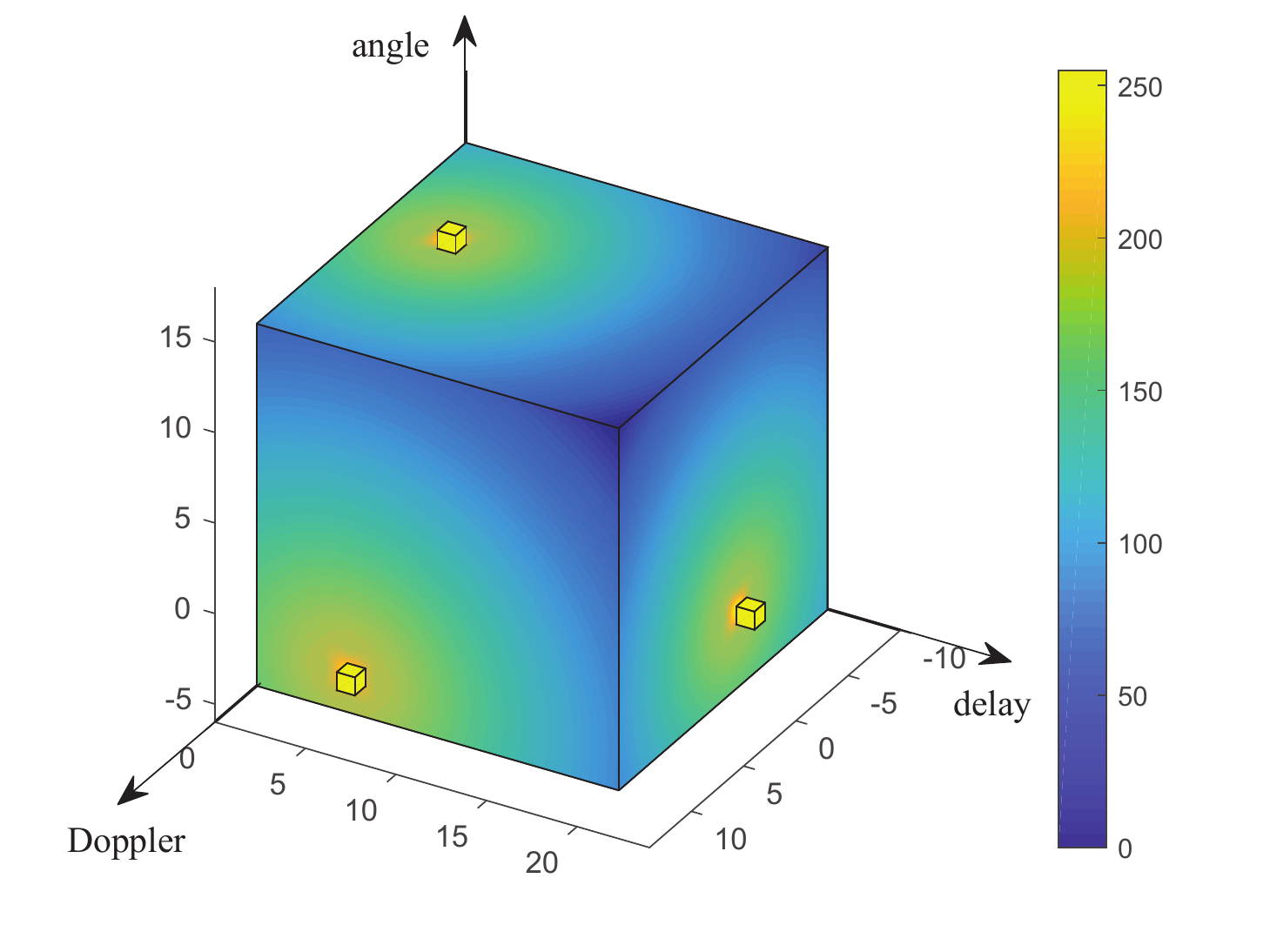}
	\caption{3D sparse channel over the delay-Doppler-angle domain.}
	\label{fig:3Dchannel}
\end{figure}

From the above equation, it can be checked that $\bar{h}_{k,i,j,q}$ has dominant elements only if $i\approx\tau_{k,p}L_D\triangle f$, $j\approx\nu_{k,p}N_DT$ and $q\approx M\frac{d\sin\theta_{k,p}}{\lambda}$, and each dominant element corresponds to one specific parameter set
$\{\tau_{k,p}, \nu_{k,p}, \theta_{k,p},
h_{k,p}\}$. Thus, there are only $P$ dominant values  among $ML_DN_D$ elements,
which means that  $\bar{h}_{k,i,j,q}$ is sparse over
the delay-Doppler-angle domain as shown in Fig. \ref{fig:3Dchannel}.

From (\ref{eq:fiveA Y^DD}), it can be checked that $x_{i,j,m}$ experiences roughly constant channel $\tilde{h}_{k,i,j,m}$ in the delay-Doppler domain and is affected
by the equivalent channel $\tilde{h}_{k,i,j,m}$.
Moreover, from (\ref{eq:fiveB H^DDA}), it can be found that $\tilde{h}_{k,i,j,m}$
can be equivalently expressed by the sparse delay-Doppler-angle domain channel $\bar{h}_{k,i,j,q}$, which
can be determined by the $P$ parameter sets $\{\tau_{k,p}, \nu_{k,p}, \theta_{k,p},
h_{k,p}\}_{p=1}^P$. Thus, once we achieve
the accurate information about the parameter sets, we can infer $\bar{h}_{k,i,j,q}$.
Theoretically,
we can achieve $\bar{h}_{k,i,j,q}$ in two ways: One way happens in the time-frequency domain to achieve the $P$ parameter sets,
while the other way works in the delay-Doppler-angle domain to directly recover $\bar{h}_{k,i,j,q}$.
However, directly recovering $\bar{h}_{k,i,j,q}$ from DL needs powerful sparse signal recovery methods with high computation burden and large training overhead, which is because the non-zero elements among $\bar{h}_{k,i,j,q}$ are unknown. 
{Thus, in the following, we propose UL-aided DL channel estimation framework, which fully exploits
the angle, the delay, and the Doppler reciprocity between UL and DL, even in the FDD mode.}

\section{SBL-based Channel Parameter Capturing along UL}

Along UL, different users separately send short training sequences to BS, and BS
captures the  $P$ parameter sets $\{\tau_{k,p}^{ul}, \nu_{k,p}^{ul}, \theta_{k,p}^{ul},
h_{k,p}^{ul}\}_{p=1}^P$ for different users in the time-frequency domain, where the superscript ``{\it ul}" denotes UL variables. Secondly, at BS we try to construct the corresponding DL parameter sets and the related $\bar{h}_{k,i,j,q}$.
Meanwhile, we can acquire the exact locations of $P$ nonzero elements in $\bar{h}_{k,i,j,q}$, and send low-overhead training to estimate the $P$ nonzero delay-Doppler-angle domain channels
at different users.

\subsection{UL Transmission Model}
{With the $P$ parameter sets $\{\tau_{k,p}^{ul}, \nu_{k,p}^{ul}, \theta_{k,p}^{ul},h_{k,p}^{ul}\}_{p=1}^P$,}
we can define UL channel of the user $k$~as
\begin{align}
\mathbf h_{k,l}^{ul}(r)=\sum_{p=1}^P {h}_{k,p}^{ul}e^{\jmath2\pi\nu^{ul}_{k,p}rT_s} \delta(lT_s-\tau_{k,p}^{ul})
\mathbf a^{ul}(\theta_{k,p}^{ul}), \label{eq: h_ul_time}
\end{align}
where $\mathbf a^{ul}(\theta_{k,p}^{ul})=\bigg[1,e^{\jmath 2\pi \frac{d\sin(\theta_{k,p}^{ul})}{\lambda^{ul}}},\ldots,e^{\jmath 2\pi (M-1) \frac{d\sin(\theta_{k,p}^{ul})}{\lambda^{ul}}}\bigg]^T$ is spatial steering vector corresponding to BS antenna array and $\lambda^{ul}$ is the carrier wavelength for the UL.

\begin{figure}[!t]
	\centering
	\includegraphics[width=3in]{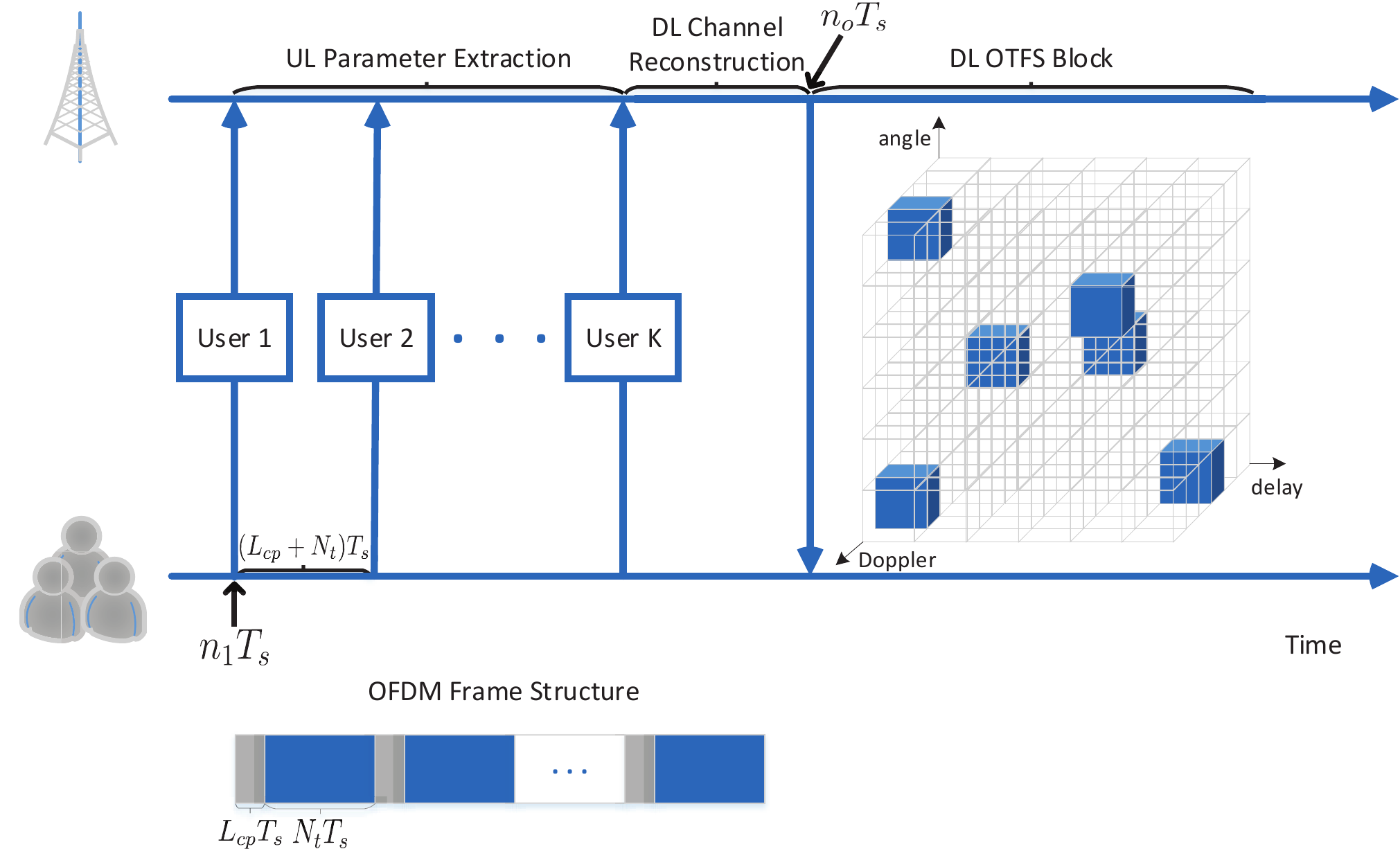}
	\caption{Overall framework in the massive MIMO-OTFS transceiver.}
	\label{fig:Overall structure}
\end{figure}

During the parameter extraction along UL,
each user sends short training sequence to BS.
The sequences from different users are orthogonal in the time domain.
As shown in Fig. \ref{fig:Overall structure}, the starting time of the training sequence for user 1 is $n_1T_s$ and the length of each training is $(L_{cp}+N_t)T_s$,
where $n_1$ is one integer number, $L_{cp}$ is the length of CP and $N_t$ is the number of valid points. The beginning time for the training of the user $k$ is $\left(n_1+(L_{cp}+N_t)(k-1)\right)T_s$, $k=1,2,\ldots,K$.
Without loss of generality, we assume that each user utilizes the same training
{$\mathbf t_{cp}=[t_{N_t-L_{cp}},t_{N_t-L_{cp}+1},\dots,t_{N_t-1},\mathbf{t}^T]^T\in\mathcal C^{(N_t+L_{cp})\times 1}$} along the UL, where {$\mathbf t=[t_0,t_1,\ldots,t_{N_t-1}]^T$}, and
$\sigma_p^2=\|\mathbf t\|^2$ represents the training power.
Then, BS receives the training sequences of the user $k$ within the time interval~$[\left(n_1+(L_{cp}+N_t)(k-1)\right)T_s,$ $\left(n_1+(L_{cp}+N_t)k-1\right)T_s]$.
It can be checked that the output of the $M$ antennas at the first $L_{cp}$ time samples in this time interval corresponds to the CP part. Hence, we will discard these samples and collect
the output of BS's antennas at time $\left(n_1+(L_{cp}+N_t)(k-1)+L_{cp}+n\right)T_s$
into the $M\times 1$ vector $\mathbf y_{k,n}^{ul}$, {$n=0,1,\ldots,N_t-1$}.

With (\ref{eq: h_ul_time}), the received vector $\mathbf y_{k,n}^{ul}$ can be denoted as
\begin{align}
\mathbf y_{k,n}^{ul}=&\sum_{p=1}^P {h}_{k,p}^{ul}e^{\jmath2\pi\nu^{ul}_{k,p}\left(n_1+(L_{cp}+N_t)(k-1)+L_{cp}+n\right)T_s} t_{(n-\tau_{k,p}^{ul}/T_s)_{N_t}}
\mathbf a^{ul}(\theta_{k,p}^{ul})+\mathbf v_{k,n},\label{eq:r_vector}
\end{align}
where $M\times 1$ vector $\mathbf v_{k,n}$ represents the additive white Gaussian noise (AWGN) vector at the $M$ receiving antennas with zero mean and
the covariance matrix $\sigma^2\mathbf I_M$. Moreover, $\mathbf v_{k,n_1}$ is independent on $\mathbf v_{k,n_2}$, when $n_1\neq n_2$.

Then, we construct the uniform sampling grids {$\{\vartheta_0,\vartheta_1,\cdots,\vartheta_{N-1}\}$} and
$[0,T_s,\ldots,(L-1)T_s]$ over the angle and delay domain, respectively, where $N$ and $L$ are the number of the angle domain and that of delay domain grids, respectively. Notice that $L$ should be greater than $\tau_{k,p}^{ul}/T_s$.
With this assumption, we can construct one $N\times L$ sparse matrix $\mathbf{G}_k^{ul}$
to map $P$ nonzero channel gains $\{h_{k,p}^{ul}\}_{p=1}^P$ onto $N$ angle and $L$ delay grids.
If we know the exact information about the $\{\tau_{k,p}^{ul}, \nu_{k,p}^{ul}, \theta_{k,p}^{ul},h_{k,p}^{ul}\}_{p=1}^P$,
we can determine the locations of the nonzero entries in $\mathbf{G}_k^{ul}$ and their values.
For example, if $\{\tau_{k,1}^{ul}, \nu_{k,1}^{ul}, \theta_{k,1}^{ul},h_{k,1}^{ul}\}$ is available,
we find the nearest point  with $\tau_{k,1}^{ul}$ among all the delay grids and label its corresponding grid as
$l_1^{*}T_s$. Similarly, the closest point with $\theta_{k,1}^{ul}$ is chosen among all the angle grids and is denoted as
$\vartheta_{n_1^*}$. Then, it can be concluded that $[\mathbf{G}_k^{ul}]_{n_1^*,l_1^{*}}=h_{k,1}^{ul}$. On
the contrary, if all the nonzero elements in $\mathbf{G}_k^{ul}$ is found, we can obtain
their related channel parameter sets. For example, if $[\mathbf{G}_k^{ul}]_{n_1^*,l_1^{*}}$ is nonzero, we have
$\tau_{k,1}^{ul}=l_1^{*}T_s$, $\theta_{k,1}=\vartheta_{n_1^*}$.
Moreover, for user $k$, the grid $[0,T_s,\ldots, (L-1)T_s]$ corresponds to  $L\times 1$ Doppler shift vector
{$\boldsymbol\upsilon_k^{ul}=[\upsilon_{k,0}^{ul},\upsilon_{k,1}^{ul},\ldots,\upsilon_{k,L-1}^{ul}]^T$}, $k=1,2,\ldots,K$.

Before proceeding, we define $N_t\times 1$ vector ${\bf{\text{exp}}}(\upsilon_{k,l}^{ul})$ and  $L\times 1$ vector $\boldsymbol{\text{exp}}_{in}(\boldsymbol{\upsilon}_k^{ul})$ as
\begin{align}
&{\bf{\text{exp}}}(\upsilon_{k,l}^{ul}) = [1,e^{\jmath2\pi\upsilon_{k,l}^{ul}T_s},\ldots,e^{\jmath2\pi\upsilon_{k,l}^{ul}(N_t-1)T_s}]^T, \notag \\
&\boldsymbol{\text{exp}}_{in}(\boldsymbol{\upsilon}_k^{ul}) = [e^{\jmath2\pi\upsilon_{k,0}^{ul} (n_1+(L_{cp}+N_t)(k-1)+L_{cp} )T_s},\ldots, e^{\jmath2\pi\upsilon_{k,L-1}^{ul} (n_1+(L_{cp}+N_t)(k-1)+L_{cp} )T_s}]^T,
\end{align}
where $l\!\!=\!\!0,1,\ldots,L\!-\!1$.
Then, we collect $\mathbf y_{k,0}^{ul},\ldots,\mathbf y_{k,N_t\!-\!1}^{ul}$ into
the matrix $\mathbf Y_{k}^{ul}\!\!=\!\![\mathbf y_{k,0}^{ul},\ldots,\mathbf y_{k,N_t\!-\!1}^{ul}]^T$.
With $\mathbf{G}_k^{ul}$ and taking some mathematical operations in \eqref{eq:r_vector}, we can denote
$\mathbf Y_{k}^{ul}$ as
\begin{align}
\mathbf{Y}_k^{ul}
=&
\underbrace{\left[\mathbf{a}^{ul}(\vartheta_0)\cdots\mathbf{a}^{ul}(\vartheta_{N-1})\right]}_{\mathbf A^{ul}}\underbrace{\mathbf G_k^{ul}\text{diag}(\boldsymbol{\text{exp}}_{in}(\boldsymbol{\upsilon}_k^{ul}))}_{\tilde{\mathbf{G}}_k^{ul}} \left[
 \begin{matrix}
   \boldsymbol{\text{exp}}^T(\upsilon_{k,0}^{ul}) \odot\left(\mathbf{t}^T\mathbf{J}_0\right)  \\
   \vdots  \\
   \boldsymbol{\text{exp}}^T(\upsilon_{k,L-1}^{ul}) \odot\left(\mathbf{t}^T\mathbf{J}_{L-1}\right)
  \end{matrix}
  \right]+\mathbf V_ k,\label{eq:R_matrix}
\end{align}
where $\mathbf{J}_l\in \mathbb{C}^{N_t\times N_t}$ is $l$ cyclic shift matrix with first column as the canonical basis vector $\mathbf{e}^{l}_{N_t}\!\!=\!\![I_{N_t}]_{:,(-l)_{N_t}}$,
and the $M\!\!\times \!\!N_t$  noise matrix $\mathbf V_k\!\!=\!\![\mathbf v_{k,0},\mathbf v_{k,1},\ldots,\mathbf v_{k,N_t-1}]$. Moreover, the $M\!\times\! N$ dictionary matrix $\mathbf A^{ul}$ and the equivalent sparse matrix $\tilde{\mathbf{G}}_k^{ul}$ have been given in
the above equation.

Nonetheless, in the practical system, the Doppler shift is much smaller than the sampling rate $1/T_s$. For example,
for the system with carrier frequency $6$ GHz, $N_t=40$ and $1/T_s=20$ MHz,  when the user moves at the speed 300 km/h, the maximum phase accumulation within $N_tT_s$ interval is 0.021, which is much smaller than 1. Thus, we can
utilize the Taylor series expansion to approximate ${\bf{\text{exp}}}(\upsilon_{k,l}^{ul})$ as
\begin{align}
{\bf{\text{exp}}}(\upsilon_{k,l}^{ul})\approx \underbrace{\left[1,1+\jmath2\pi T_s\upsilon_{k,l}^{ul},\dots,1+\jmath2\pi (N_t-1)T_s\upsilon_{k,l}^{ul}\right]^T}_{{\bf{\text{exp}}}_{ap}(\upsilon_{k,l}^{ul})},
\end{align}
where the $N_t\times 1$ vector ${\bf{\text{exp}}}_{ap}(\upsilon_{k,l}^{ul})$ is defined. Plugging the above result into \eqref{eq:R_matrix}, we can obtain

\begin{align}
\mathbf{Y}_k^{ul}
&=\mathbf A^{ul}\tilde{\mathbf{G}}_k^{ul}
 \underbrace{\left[
 \begin{matrix}
   {\bf{\text{exp}}}_{ap}^T(\upsilon_{k,0}^{ul}) \odot\left(\mathbf{t}^T\mathbf{J}_0\right)  \\
   \vdots  \\
   {\bf{\text{exp}}}_{ap}^T(\upsilon_{k,L-1}^{ul}) \odot\left(\mathbf{t}^T\mathbf{J}_{L-1}\right)
  \end{matrix}
  \right]}_{\mathbf C_k^{ul}}+\mathbf V_k,\label{eq:Y_k}
\end{align}
where the $L\times N_t$ matrix $\mathbf{C}_k^{ul}$ is denoted in the above equation.

As presented in (\ref{eq:R_matrix}), the dictionary $\mathbf{A}^{ul}$ is constructed with the uniform  angle grids $\{\vartheta_0,\dots,\vartheta_{N-1}\}$.
Correspondingly, the true DOA set for the $P$ UL scattering pathes is $\{{\theta}_{k,1}^{ul},\dots,\theta_{k,P}^{ul}\}$.
In practice, the DOAs may not locate exactly on the predefined spatial grids, and the direction mismatch happens. Under such case,
we can approximate the practical steering vector $\mathbf a^{ul}(\theta_{k,p}^{ul})$ with the linear expansion as
\begin{align}
\mathbf{a}^{ul}({{\theta}_{k,i}^{ul}})\approx\mathbf{a}^{ul}(\vartheta_{n_i})+\mathbf{b}^{ul}({\vartheta}_{n_i})({\theta}_{k,i}^{ul}-\vartheta_{n_i}),\label{eq:steer_app}
\end{align}
where  $\vartheta_{n_i}$ is the nearest angle {grid to the true DOA} $\theta_{k,i}^{ul}$,  $\mathbf{a}^{ul}(\vartheta_{n_i})$ is the steering vector on grid point $\vartheta_{n_i}$ and $\mathbf{b}^{ul}({\vartheta}_{n_i})$ is derivative of $\mathbf{a}^{ul}(\vartheta_{n_i})$ with respect to $\vartheta_{n_i}$, i.e., $\mathbf{b}^{ul}({\vartheta}_{n_i})=\left(\mathbf{a}^{ul}(\vartheta_{n_i})\right)'$.

With the off-grid taken into consideration, (\ref{eq:Y_k})  can be rewritten as
\begin{align}
\mathbf{Y}_k^{ul} =
\underbrace{\left(\mathbf{A}^{ul}+\mathbf{B}^{ul}\text{diag}(\boldsymbol{\beta}_k^{ul})\right)}_{\tilde{\mathbf A}_k^{ul}}\tilde{\mathbf{G}}_k^{ul}\mathbf{C}_k^{ul}+{\mathbf{V}}_k,
\label{eq:Y_k_off}
\end{align}
where $\mathbf{B}^{ul}=[\mathbf{b}^{ul}({\vartheta}_{0}),\dots,\mathbf{b}^{ul}({\vartheta}_{N-1})]$, $\boldsymbol{\beta}_k^{ul}=[\beta_{k,0}^{ul},\dots,\beta_{k,N-1}^{ul}]^T$,
and the $M\times N$  matrix $\tilde{\mathbf A}_k^{ul}$ is given in the above equation. Moreover,
it can be assumed that  the elements in $ \boldsymbol\beta_{k}^{ul}$ are independent identical distribution (i.i.d.) according to the uniform distribution within the region $[-\frac{r}{2},\frac{r}{2}] $ , where $ r $ is the grid interval for the uniform grid set $ \vartheta $, i.e., $ r =\vartheta_{n}-\vartheta_{n-1}, 1\leq n\leq N-1 $.

After constructing the observation model (\ref{eq:Y_k_off}), we turn the recovering of the $P$  parameter sets
$\{\tau_{k,p}^{ul}, \nu_{k,p}^{ul}, \theta_{k,p}^{ul},h_{k,p}^{ul}\}_{p=1}^P$  into  the estimation of the sparse matrix  $\tilde{\mathbf{G}}_k^{ul}$,
the basis vector $\boldsymbol\beta_k^{ul}$,  and the corresponding  Doppler shift vector {$\boldsymbol \upsilon_k^{ul}=[\upsilon_{k,0}^{ul},\upsilon_{k,1}^{ul},\ldots,\upsilon_{k,L-1}^{ul}]^T$}.
{Then, the  UL channel parameter extraction can be treated as sparse recovery problem, where
the SBL framework can achieve the robust  result \cite{whySBL}. Therefore, we will adopt a SBL framework to implement this task in the next subsection.}

\subsection{Probabilistic Models of the SBL Framework}

In the following, we will give the detailed probabilistic models for our problem. From (\ref{eq:Y_k_off}),  we can obtain  $\mathbf y^{ul}_{k,n}$ as
\begin{align}
\mathbf{y}^{ul}_{k,n}=\tilde{\mathbf{A}}_k^{ul}\tilde{\mathbf{G}}_k^{ul}\left[\mathbf{C}_k^{ul}\right]_{:,n}+{\mathbf v}_{k,n}.
\label{eq:y_k_n}
\end{align}
Hence, with the given $\boldsymbol\beta_k^{ul}$, $\boldsymbol\upsilon_k^{ul}$,
the conditional probability distribution function (PDF) of $\mathbf{y}^{ul}_{k,n}$ on $\tilde{\mathbf{G}}_k^{ul}$
can be written as
\begin{align}
p\left(\mathbf{y}^{ul}_{k,n}| \tilde{\mathbf{G}}_k^{ul};\boldsymbol{\beta}_k^{ul},\boldsymbol{\upsilon}_k^{ul}\right) = \mathcal{CN} \left(\tilde{\mathbf{A}}_k^{ul}\tilde{\mathbf{G}}_k^{ul} \left[\mathbf{C}_k^{ul}\right]_{:,n},\sigma^2\mathbf{I}_{M}\right).
\label{eq:pdf_y_kn}
\end{align}



Before proceeding, let us define the $NL\times 1$ vector $\mathbf{g}_k^{ul}=\text{vec}(\tilde{\mathbf{G}}_k^{ul})$.
With the property equation $\tilde{\mathbf{A}}_k^{ul}\tilde{\mathbf{G}}_k^{ul}[\mathbf{C}^{ul}_k]_{:,n}=(\left[\mathbf{C}_k^{ul}\right]_{:,n}^T\otimes\tilde{\mathbf{A}}_k^{ul})\text{vec}(\tilde{\mathbf{G}}_k^{ul})$ , (\ref{eq:y_k_n}) can be reformulated as
\begin{align}
\mathbf{y}^{ul}_{k,n}=\boldsymbol{\Phi}_{k,n}\mathbf{g}_k^{ul}+{\mathbf{v}}_{k,n},
\end{align}
where $\boldsymbol{\Phi}_{k,n}=\left[\mathbf{C}_k^{ul}\right]_{:,n}^T\otimes \tilde{\mathbf{A}}_k^{ul}\in\mathcal{C}^{M\times NL}$.
Let us collect
$\mathbf y^{ul}_{k,0},\mathbf y^{ul}_{k,1},\ldots,\mathbf y^{ul}_{k,N_t-1}$ into the $N_tM\times 1$ vector
$\mathbf y^{ul}_k=[(\mathbf{y}^{ul}_{k,0})^T,(\mathbf{y}^{ul}_{k,1})^T,\ldots,(\mathbf{y}^{ul}_{k,N_t-1})^T]^T$  and construct the  $N_tM \times NL$ matrix $\boldsymbol{\Phi}_{k}=\left[(\boldsymbol{\Phi}_{k,0})^T,\cdots,(\boldsymbol{\Phi}_{k,N_t\!-\!1})^T\right]^T$ and the $N_tM\!\!\times\!\! 1$ vector ${\mathbf{v}}_k\!\!=\!\!\left[{\mathbf{v}}_{k,0}^T,\cdots,{\mathbf{v}}_{k,N_t\!-\!1}^T\right]^T$.
Then, we can obtain
\begin{align}
\mathbf{y}^{ul}_{k}=\boldsymbol{\Phi}_{k}\mathbf{g}_k^{ul}+{\mathbf{v}}_k.
\end{align}

With (\ref{eq:pdf_y_kn}),  it can be obtained that
\begin{align}
p(\mathbf{y}^{ul}_{k}\mid\mathbf{g}_k^{ul};\boldsymbol{\beta}_k^{ul},\boldsymbol{\upsilon}_k^{ul})=\mathcal{CN}\left(\boldsymbol{\Phi}_{k}\mathbf{g}_k^{ul},\sigma^2\mathbf{I}_{N_tM}\right).\label{eq:threeB3 p(y_p)}
\end{align}

\begin{figure}[!t]
	\centering
	\includegraphics[width=2.5in]{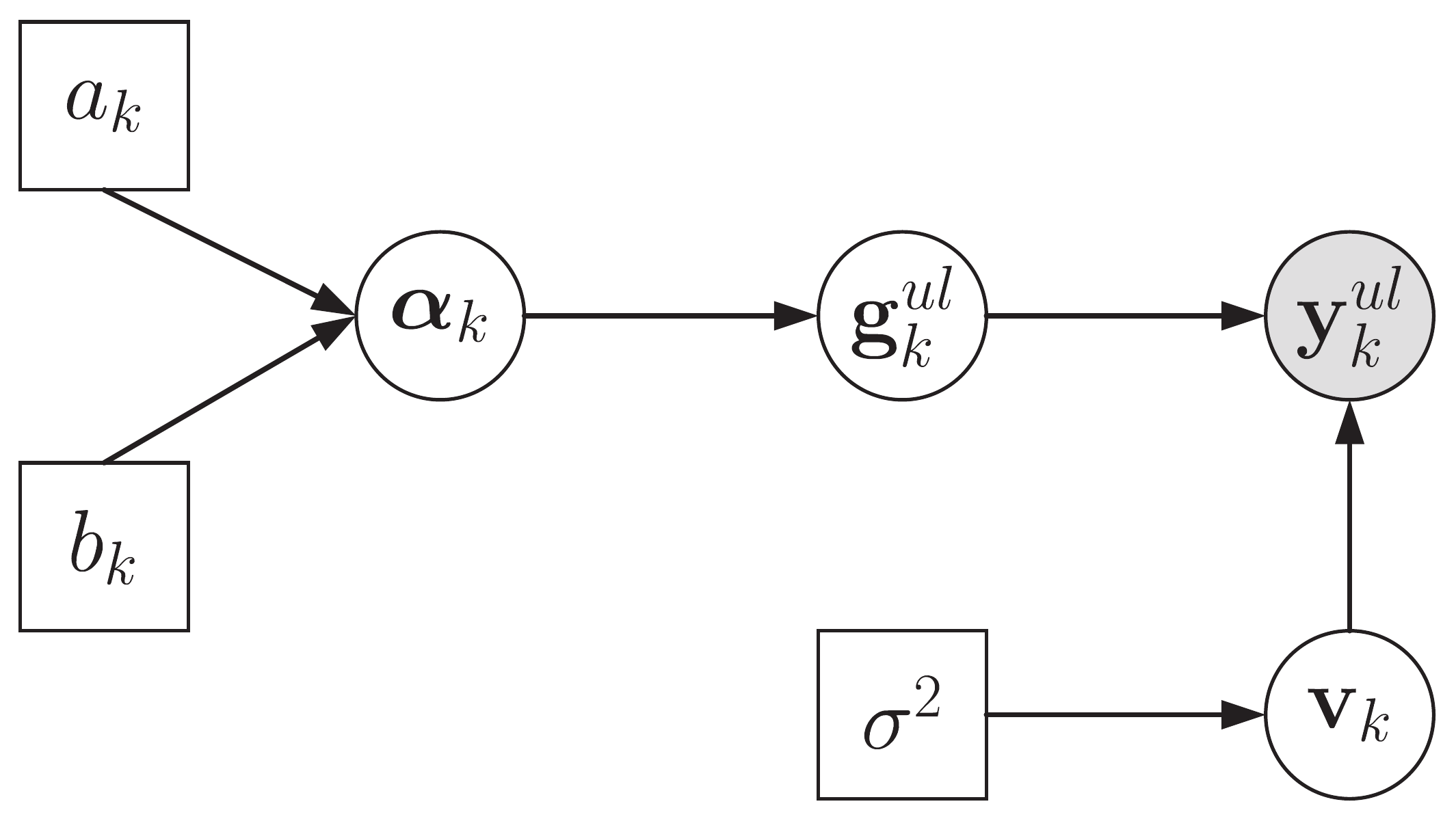}
	\caption{Graphical model for UL channel estimation using SBL.}
	\label{fig:Graphical_model}
\end{figure}

{Within the SBL framework \cite{SBL}, $\mathbf{g}_k^{ul}$ can be hierarchically modeled in Fig. \ref{fig:Graphical_model},
where the squares denote constant variables, the circles represent the hidden variables and the shaded circles correspond to the observations.}
Here, we set the prior probability of $\mathbf{g}_k^{ul}$ as the complex Gaussian distribution with precision matrix $\boldsymbol{\Gamma}_k=\text{diag}(\boldsymbol{\alpha_k})$, where $\boldsymbol{\alpha}_k=\left[\alpha_{k,0},\cdots,\alpha_{k,NL-1}\right]^T$.
 Then, the conditional PDF of $\mathbf{g}_k^{ul}$ on $\boldsymbol \alpha_k$ can be written as
 \begin{align}
p(\mathbf{g}_k^{ul}\mid\boldsymbol{\alpha}_k)=\mathcal{CN}\left(0,\boldsymbol{\Gamma}_k^{-1}\right).\label{eq:threeB3 p(h)}
\end{align}
 Furthermore, $\alpha_{k,i}$ is Gamma distributed as
\begin{align}
p(\alpha_{k,i})\!=\!\text{Gamma}(\alpha_{k,i}\!\mid\! a_k,b_k),~\text{for}\ i\!=\!0,\cdots,NL\!-\!1,\label{eq:threeB3 p(alpha)}
\end{align}
where $\text{Gamma}(\alpha_{k,i}|a_k,b_k)\!=\!\frac{b_k^{a_k}}{\Gamma(a_k)}\alpha_{k,i}^{a_k-1}\text{exp}\{-b_k\alpha_{k,i}\}$,
$\Gamma(a_k)$ is the Gamma function, and $a_k$ and $b_k$ are the constant parameters.
If $\alpha_{k,i}$ is Gamma distributed, $p(\alpha_{k,i})$ is the conjugate prior to the the conditional PDF $p([\mathbf g_k^{ul}]_i|\alpha_{k,i})$, which can simplify the calculation of the posterior distribution.
Besides, through integrating out the hyperparameter $\alpha_{k,i}$, the marginal distribution of $[\mathbf g_k^{ul}]_{i}$ is a Student-t distribution \cite{student-t}. With proper $a_k$ and $b_k$ in the Gamma distribution, the Student-t distribution is strongly peaked about the origin $[\mathbf g_k^{ul}]_i=0$, which assures the sparsity of $\mathbf{g}_k^{ul}$.

\subsection{Solving SBL Using EM-VB}
The goal of VB inference is to find a tractable variational distribution $q(\mathcal{H}_k)$ that closely approximates $p(\mathcal{H}_k|\mathbf{y}^{ul}_{k})$ \cite{posterior1}, \cite{posterior2}. Here, we denote all hidden variables as $\mathcal{H}_k = \{\tilde{\mathbf{G}}_k^{ul}, \boldsymbol{\alpha}_k\}$.
Within VB, $q(\mathcal{H}_{k,i})$ can be achieved as
\begin{align}
q\left(\mathcal{H}_{k,i}\right)
& \propto \exp \left\{\left\langle\ln p\left(\mathcal{H}_k, \mathbf{y}^{ul}_{k}\right)\right\rangle_{\sim q\left(\mathcal{H}_{k,i}\right)}\right\},\label{eq:threeC Q(H)}
\end{align}
where
$\left\langle\cdot\right\rangle_{\sim q\left(\mathcal{H}_{k,i}\right)}$ denotes {the} expectation {operation} with respect to all factors except $q(\mathcal{H}_{k,i})$.


{With} \eqref{eq:threeC Q(H)}, we can derive the posterior statistics for each variable in $\mathcal{H}_k = \{\tilde{\mathbf{G}}_k^{ul}, \boldsymbol{\alpha}_k\}$ through the expectation step of EM-VB and further derive $\{\boldsymbol{\beta}_k^{ul},\boldsymbol{\upsilon}_k^{ul}\}$
in the maximization step of EM-VB.

\subsection{Expectation Step of EM-VB}
In the expectation phase, we aim to find the approximate posterior distribution of $q(\mathbf{g}_k^{ul})$ and $q(\boldsymbol{\alpha}_k)$. Besides, $\hat{\boldsymbol{\beta}}_k^{ul}$ and $\hat{\boldsymbol{\upsilon}}_k^{ul}$ obtained from the maximization step are needed to evaluate the estimation of $\boldsymbol{\Phi}_k$, i.e., $\hat{\boldsymbol{\Phi}}_k$.

\subsubsection{Estimation of $q(\mathbf{g}_k^{ul})$}
{By applying \eqref{eq:threeB3 p(y_p)} , \eqref{eq:threeB3 p(h)} and \eqref{eq:threeC Q(H)}, we can obtain $q(\mathbf{g}_k^{ul})$ as follows}
\begin{align}
q(\mathbf{g}_k^{ul}) &\!\varpropto \!\exp\!\left\{\! \ln p(\mathbf{y}^{ul}_k | \mathbf{g}_k^{ul};\!\hat{\boldsymbol{\beta}}_k^{ul},\hat{\boldsymbol{\upsilon}}_k^{ul}) \!+\! \left\langle \ln p(\mathbf{g}_k^{ul} | \boldsymbol{\alpha}_k) \right\rangle_{q (\boldsymbol{\alpha}_k)} \right\}\notag\\
&\varpropto \exp\left\{-(\mathbf{g}_k^{ul}-\boldsymbol{\mu}_k)^H\boldsymbol{\Sigma}_k^{-1}(\mathbf{g}_k^{ul}-\boldsymbol{\mu}_k)\right\},\label{eq:qh}
\end{align}
where $\mathbf{g}_k^{ul}$ is complex Gaussian distributed with the mean and the covariance as
\begin{align}
&\boldsymbol{\mu}_k=\frac{1}{\sigma^2}\boldsymbol{\Sigma}_k\hat{\boldsymbol{\Phi}}_k^H\mathbf{y}^{ul}_k,\kern 15pt\boldsymbol{\Sigma}_k=\left(\frac{1}{\sigma^2}\hat{\boldsymbol{\Phi}}_k^H\hat{\boldsymbol{\Phi}}_k+\left\langle \boldsymbol{\Gamma}_k\right\rangle_{q(\boldsymbol{\alpha}_k)}\right)^{-1},\label{eq:threeD2 Sigma}
\end{align}
and $\left\langle\mathbf{g}_k^{ul}\right\rangle_{q(\mathbf{g}_k^{ul})}=\boldsymbol{\mu}_k$ can be used for deriving $q(\boldsymbol{\alpha}_k)$. From \eqref{eq:threeD2 Sigma}, we find that
in order to obtain $\left\langle\mathbf{g}_k^{ul}\right\rangle_{q(\mathbf{g}_k^{ul})}$, we need to derive $\left\langle \boldsymbol{\Gamma}_k\right\rangle_{q(\boldsymbol{\alpha}_k)}$ as follows.


\subsubsection{Estimation of $q(\boldsymbol{\alpha}_k)$} With \eqref{eq:threeB3 p(h)} and \eqref{eq:threeB3 p(alpha)}, $q(\boldsymbol{\alpha}_k)$ can be calculated from \eqref{eq:threeC Q(H)} as
\begin{align}
q(\boldsymbol{\alpha}_k)&\varpropto \exp\left\{\left\langle \ln p(\mathbf{g}_k^{ul}\mid\boldsymbol{\alpha}_k)\right\rangle_{q(\mathbf{g}_k^{ul})}+ \ln p(\boldsymbol{\alpha}_k)\right\}\notag\\
&\varpropto\prod_{i=0}^{NL-1}\alpha_{k,i}^{a_k+1-1}\exp\!\Big\{\!-(b_k+\big\langle [\mathbf g_k^{ul}]_i^2\big\rangle_{q(\mathbf{g}_k^{ul})})\alpha_{k,i}\!\Big\}\!.\label{eq:qalpha}
\end{align}

Thus, $\alpha_{k,i}$ subjects to $\text{Gamma}(\alpha_{k,i};\widetilde{a}_k,\widetilde{b}_k)$, where $\widetilde{a}_k=a_k+1$ and $\widetilde{b}_k=b_k+\left\langle \left[\mathbf g_k^{ul}\right]_i^2\right\rangle_{q(\mathbf{g}_k^{ul})}$. From the property of the Gamma distribution, we can obtain
\begin{align}
\left\langle \alpha_{k,i}\right\rangle_{q(\boldsymbol{\alpha}_k)}=\frac{\widetilde{a}_k}{\widetilde{b}_k}=\frac{a_k+1}{b_k+\left\langle \left[\mathbf g_k^{ul}\right]_i^2\right\rangle_{q(\mathbf{g}_k^{ul})}},\label{eq:threeD alpha}
\end{align}
where $a_k$ and $b_k$ can be designed as a reasonable value, and
$\left\langle \left[\mathbf g_k^{ul}\right]_i^2\right\rangle_{q(\mathbf{g}_k^{ul})}$ is the $i$-th element of $\text{diag}(\boldsymbol{\mu}_k\boldsymbol{\mu}_k^H+\boldsymbol{\Sigma}_k)$.
Finally, we can obtain
\begin{align}
\left\langle \boldsymbol{\Gamma}_k\right\rangle_{q(\boldsymbol{\alpha}_k)}
\!=\!\text{diag}\!\Big(\!\big[\left\langle \alpha_{k,0}\right\rangle_{q(\boldsymbol{\alpha}_k)}\!,\!\ldots\!,\!\left\langle \alpha_{k,NL-1}\right\rangle_{q(\boldsymbol{\alpha}_k)}\!\big]^T\!\Big),
\end{align}
which is needed in deriving $\left\langle\mathbf{g}_k^{ul}\right\rangle_{q(\mathbf{g}_k^{ul})}$.

\subsection{Maximization Step of EM-VB}

With \eqref{eq:threeD2 Sigma} and the definitions of both $\tilde{\mathbf{G}}_k^{ul}$ and $\mathbf g^{ul}_k$, we can achieve $\langle\tilde{\mathbf{G}}_k^{ul}\rangle_{q(\mathbf{g}_k^{ul})}$ as $\langle\tilde{\mathbf{G}}_k^{ul}\rangle_{q(\mathbf{g}_k^{ul})}=[[\boldsymbol{\mu}_k]_{0:N-1},\cdots,[\boldsymbol{\mu}_k]_{(N-1)L:NL-1}]$, which is needed in the derivation of both $\boldsymbol{\beta}_k^{ul}$ and $\boldsymbol{\upsilon}_k^{ul}$.
\subsubsection{Computing of $\boldsymbol{\beta}_k^{ul}$}In the maximization step of the EM-VB, $\boldsymbol{\beta}_k^{ul}$
can be estimated as
\begin{align}
\hat{\boldsymbol{\beta}}_k^{ul}=\max\limits_{\text{diag}(\boldsymbol{\beta}_k^{ul})}\left\langle
\ln p(\mathbf{Y}_k^{ul}\mid\tilde{\mathbf{G}}_k^{ul};\boldsymbol{\beta}_k^{ul},\boldsymbol{\upsilon}_k^{ul})\right\rangle_{q(\mathbf{g}_k^{ul})},
\end{align}
which is equivalent to \eqref{beta_2} as follows
\begin{align}
&\hat{\boldsymbol{\beta}}_k^{ul} = \min\limits_{\boldsymbol{\beta}_k^{ul}} \left\langle \bigg\{ \sum_{n=0}^{N_t-1} \frac{1}{\sigma^2} \left[\mathbf{y}_{k,n}^{ul} - \mathbf{A}^{ul} \tilde{\mathbf{G}}_k^{ul} [\mathbf{C}_k^{ul}]_{:,n} - \mathbf{B}^{ul}\text{diag} (\tilde{\mathbf{G}}_k^{ul} [\mathbf{C}_k^{ul}]_{:,n}) \boldsymbol{\beta}_k^{ul}\right]^H\right.\notag\\
&\kern 105pt \left.\cdot \left[\mathbf{y}_{k,n}^{ul} - \mathbf{A}^{ul} \tilde{\mathbf{G}}_k^{ul} [\mathbf{C}_k^{ul}]_{:,n} -\mathbf{B}^{ul} \text{diag}(\tilde{\mathbf{G}}_k^{ul} [\mathbf{C}_k^{ul}]_{:,n}) \boldsymbol{\beta}_k^{ul}\right] \bigg\}\right\rangle_{q(\mathbf{g}_k^{ul})} \label{beta_2}
\end{align}

Taking the derivation of the above equation and let it equal to zero with respect to $\boldsymbol{\beta}_k^{ul}$, we have
\begin{align}
&\hat{\boldsymbol{\beta}}_k^{ul}=
\left[\sum_{n=0}^{N_t-1} \Re\left\{ \text{diag} \left(\left( \left\langle\tilde{\mathbf{G}}_k^{ul} \right\rangle_{q(\mathbf{g}_k^{ul})} [\mathbf{C}_k^{ul}]_{:,n}\right)^H \right) (\mathbf{B}^{ul})^H \mathbf{B}^{ul} \text{diag} \left(\left\langle \tilde{\mathbf{G}}_k^{ul} \right\rangle_{q(\mathbf{g}_k^{ul})} [\mathbf{C}_k^{ul}]_{:,n}\right) \right\}\right]^{-1}\notag\\
&\kern 4pt \cdot\sum_{n=0}^{N_t-1} \Re\left\{\text{diag} \left(\left(\left\langle \tilde{\mathbf{G}}_k^{ul} \right\rangle_{q(\mathbf{g}_k^{ul})} [\mathbf{C}_k^{ul}]_{:,n}\right)^H\right) (\mathbf{B}^{ul})^H \left(\mathbf{y}_{k,n}^{ul} - \mathbf{A}^{ul} \left\langle\tilde{\mathbf{G}}_k^{ul} \right\rangle_{q(\mathbf{g}_k^{ul})} [\mathbf{C}_k^{ul}]_{:,n}\right) \right\}.\label{eq:threeF beta}
\end{align}

\subsubsection{Computing of $\boldsymbol{\upsilon}_k^{ul}$}
Let us define the $L\times N_t$  matrix $\mathbf{D}_k=\left[(\mathbf{t}^T\mathbf{J}_0)^T,\cdots,(\mathbf{t}^T\mathbf{J}_{L-1})^T\right]^T$. Then, we can obtain $\mathbf{C}_k^{ul}=\left[{\bf{\text{exp}}}_{ap}(\upsilon_{k,0}^{ul}),\ldots,{\bf{\text{exp}}}_{ap}(\upsilon_{k,L-1}^{ul})\right]^T\odot\mathbf{D}_k$ and $[\mathbf{C}_k^{ul}]_{:,n}=\text{diag}([\mathbf{D}_k]_{:,n})\times\left(\mathbf{1}_L+\jmath2\pi nT_s\boldsymbol{\upsilon}_k^{ul}\right)$.

Then the estimations of $\boldsymbol{\upsilon}_k^{ul}$ can be given by
\begin{align}
\hat{\boldsymbol{\upsilon}}_k^{ul}=\max\limits_{\boldsymbol{\upsilon}_k^{ul}}\left\langle \sum_{n=0}^{N_t-1} \ln p(\mathbf{y}_{k,n}^{ul}\mid\tilde{\mathbf{G}}_k^{ul};\boldsymbol{\beta}_k^{ul},\boldsymbol{\upsilon}_k^{ul})\right\rangle_{q(\mathbf{g}_k^{ul})},
\end{align}
which is equivalent to \eqref{nu_f} as follows
\begin{align}
&\hat{\boldsymbol{\upsilon}}_k^{ul}=\! \min\limits_{ \boldsymbol{\upsilon}_k^{ul}} \left\langle \!\sum_{n=0}^{N_t-1} \!\bigg\{ \!\left[ \mathbf{y}_{k,n}^{ul} \!\!-\! \tilde{\mathbf{A}}_k^{ul} \tilde{\mathbf{G}}_k^{ul} \text{diag} ([\mathbf{D}_k]_{:,n})\! \left(\mathbf{1}_L \!\!+ \! \jmath2\pi nT_s \boldsymbol{\upsilon}_k^{ul} \right) \!\right]^H \!\! \right.\notag\\
&\left.\kern 120pt\cdot\! \left[\mathbf{y}_{k,n}^{ul} \!\!-\!\! \tilde{\mathbf{A}}_k^{ul} \tilde{\mathbf{G}}_k^{ul} \text{diag} ([\mathbf{D}_k]_{:,n}) \!\left(\mathbf{1}_L \!\!+\! \jmath2\pi nT_s \boldsymbol{\upsilon}_k^{ul} \right) \!\right] \!\bigg\} \!\right\rangle_{ q(\mathbf{g}_k^{ul})}\notag\\
&\kern 15pt= \min \limits_{\boldsymbol{\upsilon}_k^{ul}} \left\langle \sum_{n=0}^{N_t-1} \bigg\{\Big[(2\pi nT_s)^2 (\boldsymbol{\upsilon}_k^{ul})^H \text{diag} ([\mathbf{D}_k]_{:,n}^H) (\tilde{\mathbf{G}}_k^{ul})^H (\tilde{\mathbf{A}}_k^{ul})^H \tilde{\mathbf{A}}_k^{ul} \tilde{\mathbf{G}}_k^{ul} \text{diag} ([\mathbf{D}_k]_{:,n})\boldsymbol{\upsilon}_k^{ul}\right.\notag\\
&\kern 120pt\left.-(\jmath2\pi nT_s) (\mathbf{O}_{k,n}^{ul})^H \tilde{\mathbf{A}}_k^{ul} \tilde{\mathbf{G}}_k^{ul} \text{diag}([\mathbf{D}_k]_{:,n})\boldsymbol{\upsilon}_k^{ul} \right.\notag\\
&\left.\kern 120pt+(\jmath2\pi nT_s) (\boldsymbol{\upsilon}_k^{ul})^H \text{diag}([\mathbf{D}_k]_{:,n}^H) (\tilde{\mathbf{G}}_k^{ul})^H (\tilde{\mathbf{A}}_k^{ul})^H \mathbf{O}_{k,n}^{ul} \Big]\bigg\} \right\rangle_{q(\mathbf{g}_k^{ul})}, \label{nu_f}
\end{align}
and $\mathbf{O}_{k,n}^{ul}=\mathbf{y}_{k,n}^{ul}- \tilde{\mathbf{A}}_k^{ul}\tilde{\mathbf{G}}_k^{ul}\text{diag}([\mathbf{D}_k]_{:,n})\mathbf{1}_L$ in \eqref{nu_f}.

With the similar operations in deriving $\hat{\boldsymbol{\beta}}_k^{ul}$, we can obtain
\begin{align}
\hat{\boldsymbol{\upsilon}}_k^{ul}=\boldsymbol{\Pi}_k^{-1}\boldsymbol{\omega}_k,\label{eq:threeF nu}
\end{align}
where
\begin{align}
\boldsymbol{\Pi}_k=&\sum_{n=0}^{N_t-1}(2\pi nT_s)\Re\left\{\text{diag}([\mathbf{D}_k]_{:,n}^H) \left\langle\tilde{\mathbf{G}}_k^{ul} \right\rangle_{q(\mathbf{g}_k^{ul})}^H (\tilde{\mathbf{A}}_k^{ul})^H\tilde{\mathbf{A}}_k^{ul} \left\langle\tilde{\mathbf{G}}_k^{ul}\right\rangle_{q(\mathbf{g}_k^{ul})} \text{diag}([\mathbf{D}_k]_{:,n})\right\},\notag
\end{align}
and
\begin{align}
\boldsymbol{\omega}_k=&\sum_{n=0}^{N_t-1} \Im\left\{\text{diag}([\mathbf{D}_k]_{:,n}^H) \left\langle\tilde{\mathbf{G}}_k^{ul}\right\rangle_{q(\mathbf{g}_k^{ul})}^H (\tilde{\mathbf{A}}_k^{ul})^H \!\left(\!\mathbf{y}_{k,n}^{ul}\!-\!\tilde{\mathbf{A}}_k^{ul} \left\langle\tilde{\mathbf{G}}_k^{ul}\right\rangle_{q(\mathbf{g}_k^{ul})} \text{diag}([\mathbf{D}_k]_{:,n})\mathbf{1}_L\!\right)\!\right\}\notag.
\end{align}

With $\{\hat{\boldsymbol{\beta}}_k^{ul},\hat{\boldsymbol{\upsilon}}_k^{ul}\}$ in the maximization step, we can update $q(\mathbf{g}_k^{ul})$ and $q(\boldsymbol{\alpha}_k)$ to approximate  $p\left(\mathbf{g}_k^{ul} | \mathbf{y}^{ul}_{k}\right)$ and $p\left(\boldsymbol{\alpha}_k | \mathbf{y}^{ul}_{k}\right)$ in \eqref{eq:qh} and \eqref{eq:qalpha}, respectively. Furthermore,
with $q(\mathbf{g}_k^{ul})$ given in the expectation step, we can separately obtain  the estimation of $\boldsymbol{\beta}_k^{ul}$ and $\boldsymbol{\upsilon}_k^{ul}$
in \eqref{eq:threeF beta}, \eqref{eq:threeF nu}  during the maximization step.
Iteratively implementing both the expectation and the maximization steps,
we can separately obtain the estimation of  $\mathbf{g}_k^{ul}$ as $\hat{\mathbf{g}}_k^{ul}=\left\langle\mathbf{g}_k^{ul}\right\rangle_{q(\mathbf{g}_k^{ul})}$.
Correspondingly, the EM-VB scheme is summarized in \textbf{Algorithm} \ref{alg:VBEM scheme}.

\begin{algorithm}
 \caption{UL parameter extraction of user $k$ with EM-VB scheme} 
 \begin{algorithmic}[1]\label{alg:VBEM scheme}
 \STATE {\bf Input:} Training vector $\mathbf t$, hyper-parameters in $\{a_k,b_k\}$.
 \STATE {\bf Initialize:} $N_{maxiter}$ and the unknown vector set $\{\boldsymbol{\beta}_k^{ul},\boldsymbol{\upsilon}_k^{ul}\}$.
  \WHILE  {$l^{EM}<N_{maxiter}$}
  \STATE $l^{EM}=1$.
  \STATE \textbf{E-step}:
       \STATE \ \ Update $\mathbf{g}_k^{ul}$ by \eqref{eq:threeD2 Sigma}.
       \STATE \ \ Update $\boldsymbol{\alpha}_k$ by \eqref{eq:threeD alpha}.
  \STATE \textbf{M-step}:
       \STATE \ \ Update $\boldsymbol{\beta}_k^{ul}$ by \eqref{eq:threeF beta}.
       \STATE \ \ Update $\boldsymbol{\upsilon}_k^{ul}$ by \eqref{eq:threeF nu}.
  \STATE $l^{EM}\gets l^{EM}+1$.
  \ENDWHILE
  \RETURN $\mathbf{g}^{ul}_k$, $\boldsymbol{\beta}_k^{ul}$ and $\boldsymbol{\upsilon}_k^{ul}$.
 \end{algorithmic}
\end{algorithm}

\subsection{Low Complex EM-VB}
The EM-VB in the previous subsection is mathematically reasonable and efficient.
However, it can be seen from \eqref{eq:threeD2 Sigma}
that EM-VB  requires the inversion of $NL\times NL$ matrix at each iteration step.
In order to overcome this bottleneck, we will  resort to the fast Bayesian inference to design the low complex EM-VB, which also contains the expectation and the maximization steps \cite{fastVB}.
Explicitly, within the expectation step, $\boldsymbol{\alpha}_k$ can be recovered through the
maximum a posteriori (MAP) estimator as
\begin{align}
\hat{\boldsymbol\alpha}_k=\max\limits_{{\boldsymbol\alpha}_k}\mathcal{L}(\boldsymbol{\alpha}_k),
\textit{s.t.}~\alpha> 0 ,\label{eq:MAP_alpha_vector}
\end{align}
where
\begin{align}
\mathcal{L}(\boldsymbol{\alpha}_k)=& \log p(\mathbf{y}_k^{ul}\mid\boldsymbol{\alpha}_k;\hat{\boldsymbol{\beta}}_k^{ul},\hat{\boldsymbol{\upsilon}}_k^{ul})\notag\\
=&\!-\!\big[MN_t\log\pi\!+\!\log|\underbrace{\sigma^2\mathbf{I}_{MN_t}
\!+\!\hat{\boldsymbol{\Phi}}_k\boldsymbol{\Gamma}^{-1} \hat{\boldsymbol{\Phi}}_k^H}_{\boldsymbol{\Xi}_k}| \!+\!(\mathbf{y}_k^{ul})^H(\sigma^2\mathbf{I}_{MN_t}\!+\!\hat{\boldsymbol{\Phi}}_k\boldsymbol{\Gamma}^{-1}\hat{\boldsymbol{\Phi}}_k^H)^{-1}\mathbf{y}_k^{ul}\big],\label{eq:Lalpha}
\end{align}
and $\boldsymbol{\Xi}_k$ is presented in the above equation.
Moreover, instead of deriving all the elements of $\boldsymbol\alpha_k$ in (\ref{eq:qalpha}) within
each expectation step of the EM-VB, only one entry in $\boldsymbol\alpha_k$ would be updated in each iteration
of the fast Bayesian inference.

Before proceeding, we give the following matrix properties.
From the matrix theory, $\boldsymbol{\Xi}_k$ in \eqref{eq:Lalpha} can be rewritten as
\begin{align}
\boldsymbol{\Xi}_k&=\sigma^2\mathbf{I}_{MN_t}+\sum_{j\neq i}\alpha_{k,j}^{-1}[\hat{\boldsymbol{\Phi}}_k]_{:,j}[\hat{\boldsymbol{\Phi}}_k]_{:,j}^H+\alpha_{k,i}^{-1}[\hat{\boldsymbol{\Phi}}_k]_{:,i}[\hat{\boldsymbol{\Phi}}_k]_{:,i}^H\notag\\
&=\boldsymbol{\Xi}_{k,-i}+\alpha_{k,i}^{-1}[\hat{\boldsymbol{\Phi}}_k]_{:,i}[\hat{\boldsymbol{\Phi}}_k]_{:,i}^H,
\end{align}
where $\boldsymbol{\Xi}_{k,-i}$ contains all items of $\boldsymbol{\Xi}_k$ without the items related to $\alpha_{k,i}$. With the matrix inversion lemma and the determinant property, we can obtain
\begin{align}
&\boldsymbol{\Xi}_k^{-1}=\boldsymbol{\Xi}_{k,-i}^{-1}- \frac{\boldsymbol{\Xi}_{k,-i}^{-1}[\hat{\boldsymbol{\Phi}}_k]_{:,i} [\hat{\boldsymbol{\Phi}}_k]_{:,i}^H\boldsymbol{\Xi}_{k,-i}^{-1}} {\alpha_{k,i}+[\hat{\boldsymbol{\Phi}}_k]_{:,i}^H\boldsymbol{\Xi}_{k,-i}^{-1} [\hat{\boldsymbol{\Phi}}_k]_{:,i}}, \notag \\
&|\boldsymbol{\Xi}_k|=|\boldsymbol{\Xi}_{k,-i}||1+\alpha_{k,i}^{-1}[\hat{\boldsymbol{\Phi}}_k]_{:,i}^H\boldsymbol{\Xi}_{k,-i}^{-1}[\hat{\boldsymbol{\Phi}}_k]_{:,i}|.\label{eq:Xideter}
\end{align}

Since only $\alpha_{k,i}$ needs to be updated with the other elements of $\boldsymbol\alpha_{k}$ unchanging, we
can decompose the $\mathcal{L}(\boldsymbol{\alpha}_k)$ as
\begin{align}
\mathcal{L}(\boldsymbol{\alpha}_k)
=&\mathcal{L}(\boldsymbol{\alpha}_{k,-i}) \!+\! \log\alpha_{k,i} \!-\! \log(\alpha_{k,i}\!+\!\underbrace{[\hat{\boldsymbol{\Phi}}_k]_{:,i}^H \boldsymbol{\Xi}_{k,-i}^{-1}[\hat{\boldsymbol{\Phi}}_k]_{:,i}}_{p_{k,i}}) + \frac{\Big(\overbrace{\|[\hat{\boldsymbol{\Phi}}_k]_{:,i}^H \boldsymbol{\Xi}_{k,-i}^{-1} \mathbf{y}_k^{ul}\|}^{q_{k,i}}\Big)^2}{\alpha_{k,i}+ [\hat{\boldsymbol{\Phi}}_k]_{:,i}^H\boldsymbol{\Xi}_{k,-i}^{-1} [\hat{\boldsymbol{\Phi}}_k]_{:,i}}\notag\\
=&\mathcal{L}(\boldsymbol{\alpha}_{k,-i})+\ell(\alpha_{k,i}),\label{eq:L_alpha}
\end{align}
where $\ell(\alpha_{k,i})=\log\alpha_{k,i}-\log(\alpha_{k,i}+p_{k,i})+\frac{q_{k,i}^2}{\alpha_{k,i}+p_{k,i}}$,
$\mathcal{L}(\boldsymbol{\alpha}_{k,-i})$ is one function of the $(NL-1)\times 1$ vector
$\boldsymbol{\alpha}_{k,-i}=[\alpha_{k,0},\ldots,\alpha_{k,i-1},\alpha_{k,i+1},\ldots,\alpha_{k,NL-1}]^T$, and the terms $p_{k,i}$ and $q_{k,i}$ are defined in the above equations.
Notice that $\mathcal{L}(\boldsymbol{\alpha}_{k,-i})$, $p_{k,i}$ and $q_{k,i}$ do not depend on $\alpha_{k,i}$.
Moreover, (\ref{eq:Xideter}) is utilized in the above derivation.

Then, with (\ref{eq:MAP_alpha_vector}) and (\ref{eq:L_alpha}), for fixed $\boldsymbol{\alpha}_{k,-i}$, $\alpha_{k,i}$ can be estimated as
\begin{align}
\hat{\alpha}_{k,i}=\max_{\alpha_{k,i}} \ell(\alpha_{k,i}), \textit{s.t.}~\alpha_{k,i}>0.
\end{align}
Taking the derivative of  $\ell(\alpha_{k,i})$ with respect to $\alpha_{k,i}$, we have
\begin{align}
\frac{d\ell(\alpha_{k,i})}{d\alpha_{k,i}}=\frac{1}{\alpha_{k,i}}-\frac{1}{\alpha_{k,i}+p_{k,i}}-\frac{q_{k,i}^2}{(\alpha_{k,i}+p_{k,i})^2}.
\end{align}

With fixed $\boldsymbol{\alpha}_{k,-i}$, $\alpha_{k,i}$ can be updated as
\begin{align}
&\alpha_{k,i}=\left\{
\begin{aligned}\frac{p_{k,i}^2}{q_{k,i}^2-p_{k,i}},\kern 45pt \text{if}\ q_{k,i}^2>p_{k,i},\notag\\
\infty,\kern 82pt \text{if}\ q_{k,i}^2\leqslant p_{k,i}.
\end{aligned}
\right.\label{eq:judgment}
\end{align}

Interestingly, $\alpha_{k,i}=\infty$ means that the variance of $[\mathbf{h}_k^{ul}]_i$ is 0, and
the column vector $[\hat{\boldsymbol{\Phi}}_k]_{:,i}$ has no contribution to the observation vector $\mathbf y_k^{ul}$.
Hence, we can prune the basis $[\hat{\boldsymbol{\Phi}}_k]_{:,i}$ out of the observation signal space,
which can effectively decrease the problem dimension.

Iteratively implementing the above steps from
$\alpha_{k,0}$ to $\alpha_{k,NL-1}$, we can obtain the effective signal observation space
 $\mathcal B=\{[\hat{\boldsymbol{\Phi}}_k]_{:,i}|\alpha_{k,i}\neq\infty\}$,
 and collect all the column vectors in  $\mathcal B$ to form the matrix $\boldsymbol{\widehat\Phi}_{k}^e$.
To clearly illustrate the updating process of $\boldsymbol\alpha_k$,
we use the superscript $(l)$ to represent the variables in the $l$-th iteration.
According to the status of both $\alpha_{k,i}^{(l-1)}$ and $\mathcal B^{(l-1)}$,
we have three operations in current $l$-th iteration.
For $\alpha_{k,i}^{(l-1)}<\infty$ and
$[q_{k,i}^{(l-1)}]^2>p^{(l-1)}_{k,i}$, which means
that $[\hat{\boldsymbol{\Phi}}_k]_{:,i}$ belongs to $\mathcal B^{(l-1)}$ within the $(l-1)$-th iteration , we have
$\mathcal B^{(l)}=\mathcal B^{(l-1)}$ and \textbf{re-estimate} $\alpha_{k,i}^{(l)}$ as
$\frac{\left[p^{(l-1)}_{k,i}\right]^2}{\left[q_{k,i}^{(l-1)}\right]^2-p^{(l-1)}_{k,i}}$.
For $\alpha_{k,i}^{(l-1)}=\infty$ and $[q_{k,i}^{(l-1)}]^2>p^{(l-1)}_{k,i}$, which corresponds
to the case that $[\hat{\boldsymbol{\Phi}}_k]_{:,i}$ does not lie in $\mathcal B^{(l-1)}$, we should \textbf{add}
$[\hat{\boldsymbol{\Phi}}_k]_{:,i}$ into $\mathcal B^{(l)}$, i.e., $\mathcal B^{(l)}=\{\mathcal B^{(l-1)},[\hat{\boldsymbol{\Phi}}_k]_{:,i}\}$, and update $\alpha_{k,i}^{(l)}$ as
$\frac{\left[p^{(l-1)}_{k,i}\right]^2}{\left[q_{k,i}^{(l-1)}\right]^2-p^{(l-1)}_{k,i}}$.
If $\alpha_{k,i}^{(l-1)}<\infty$ and
$[q_{k,i}^{(l-1)}]^2\le p^{(l-1)}_{k,i}$, we can verify that $[\hat{\boldsymbol{\Phi}}_k]_{:,i}$ lies in $\mathcal B^{(l-1)}$ but should not be put into $\mathcal B^{(l)}$. So, we \textbf{prune} $\{[\hat{\boldsymbol{\Phi}}_k]_{:,i}\}$ out of
 $\mathcal B^{(l-1)}$ to obtain
$\mathcal B^{(l)}=\mathcal B^{(l-1)}\backslash\{[\hat{\boldsymbol{\Phi}}_k]_{:,i}\}$ and set
$\alpha^{(l)}_{k,i}$ as $\infty$.

With the similar methods in \cite{fast} and (\ref{eq:L_alpha}), the term $q_{k,i}^{(l)}$
and $p^{(l)}_{k,i}$ can be achieved from the following equations
\begin{align}
p_{k,i}^{(l)}&=\frac{\alpha_{k,i}^{(l)}P_{k,i}^{(l)}}{\alpha_{k,i}^{(l)}-P_{k,i}^{(l)}},\kern 30pt
q_{k,i}^{(l)}=\frac{\alpha_{k,i}^{(l)}Q_{k,i}^{(l)}}{\alpha_{k,i}^{(l)}-P_{k,i}^{(l)}},\\ \label{eq:pq}
P_{k,i}^{(l)}&=\sigma^{-2}[\hat{\boldsymbol{\Phi}}_k]_{:,i}^H [\widehat{\boldsymbol{\Phi}}_k]_{:,i} -\sigma^{-4}  [\hat{\boldsymbol{\Phi}}_k]_{:,i}^H \widehat{\boldsymbol{\Phi}}_k^{e,(l)} \boldsymbol{\Sigma}_k^{e,(l)} (\widehat{\boldsymbol{\Phi}}_k^{e,(l)})^H [\hat{\boldsymbol{\Phi}}_k]_{:,i},\\
Q_{k,i}^{(l)}&=\sigma^{-2} [\hat{\boldsymbol{\Phi}}_k]_{:,i}^H \mathbf{y}_k^{ul}-\sigma^{-4} [\hat{\boldsymbol{\Phi}}_k]_{:,i}^H \widehat{\boldsymbol{\Phi}}_k^{e,(l)} \boldsymbol{\Sigma}_k^{e,(l)} (\widehat{\boldsymbol{\Phi}}_k^{e,(l)})^H\mathbf{y}_k^{ul},
\end{align}
where $\widehat{\boldsymbol{\Phi}}_k^{e,(l)}$ corresponds to $\mathcal B^{(l)}$.
$\boldsymbol{\mu}_k^{e,(l)}$
and $\boldsymbol{\Sigma}_k^{e,(l)}$ can be defined
from (\ref{eq:threeD2 Sigma}) as
\begin{align}
\boldsymbol{\mu}_k^{e,(l)}&=\frac{1}{\sigma^2}\boldsymbol{\Sigma}_k^{e,(l)}(\widehat{\boldsymbol{\Phi}}_k^{e,(l)})^H\mathbf{y}^{ul}_k,\label{eq:mu_eff}\\
\boldsymbol{\Sigma}_k^{e,(l)}&=\left(\frac{1}{\sigma^2}(\widehat{\boldsymbol{\Phi}}_k^{e,(l)})^H\widehat{\boldsymbol{\Phi}}_k^{e,(l)}+ \boldsymbol{\Gamma}_k^{e,(l)}\right)^{-1}, \label{eq:Sigma_eff}
\end{align}
and $\boldsymbol{\Gamma}_k^{e,(l)}$ is constructed from the diagonal matrix
$\text{diag}\big\{\alpha_{k,0}^{(l)},\alpha_{k,1}^{(l)}, \ldots, \alpha_{k,NL-1}^{(l)}\big\}$
through deleting the diagonal elements with value of $\infty$. It can be checked
that $\widehat{\boldsymbol{\Phi}}_k^{e,(l)}$, $\mathcal B^{(l)}$, and $\boldsymbol{\Sigma}_k^{e,(l)}$ only correspond
to the non-zero positions in the sparse vector $\mathbf h_k^{(ul)}$. Hence,
the inversion operation in (\ref{eq:Sigma_eff}) would be much simpler than that in (\ref{eq:threeD2 Sigma}).
Fortunately, when only one element in $\boldsymbol\alpha_k$ is updated at each iteration,
$\boldsymbol{\Sigma}_k^{e,(l)}$ can be sequentially derived from $\boldsymbol{\Sigma}_k^{e,(l-1)}$, whose details
can be found in the appendix of \cite{fast}.

Notice that the fast Bayesian inference just modify the recovering of $\boldsymbol\alpha_k$ in expectation step of EM-VB.
Correspondingly, we summarize the low complex EM-VB in~\textbf{Algorithm}~\ref{alg:Fast interfence scheme}.

\subsection{Complexity Analysis}

{For the VB method, the computational complexity mainly depends on the calculation of the matrix inverse $\boldsymbol{\Sigma}_k$ in \eqref{eq:threeD2 Sigma}, where a computational complexity in the order of $\mathcal{O}(N^3L^3)$ is required.}

{Due to the high computational complexity, we resort to the fast VB method to overcome this bottleneck. The computational complexity can be reduced to $\mathcal{O}(M^3N_t^3)$, which is due to the fact that $MN_t< NL$.
Moreover, in the fast VB method,  a single $\alpha_{k,i}$ will be updated at each iteration  instead of updating the whole $\alpha_{k,i},i\in\{0,1,\ldots,NL-1\}$, which
leads to very efficient updates of the mean $\boldsymbol{\mu}_k^{e,(l)}$ and the covariance matrix $\boldsymbol{\Sigma}_k^{e,(l)}$ in \eqref{eq:mu_eff} and \eqref{eq:Sigma_eff}, respectively. Because  $\mathbf{g}^{ul}_k$ is highly sparse, $\boldsymbol{\Sigma}_k$ can be constructed with fewer dimensions than $NL\times NL$, which further reduces the computational complexity.}

\begin{algorithm}
 \caption{UL parameter extraction of user $k$ with fast variational Bayesian procedure} 
 \begin{algorithmic}[1]\label{alg:Fast interfence scheme}
  \STATE {\bf Input:} Training vector $\mathbf t$, hyper-parameters in $\{a_k,b_k\}$.
  \STATE {\bf Initialize:} $N_{maxiter}$ and the unknown vector set $\{\boldsymbol{\beta}_k^{ul},\boldsymbol{\upsilon}_k^{ul}\}$.
  \WHILE {$l^{EM}<N_{maxiter}$}
  \STATE $l^{EM}=1$.
  \STATE \textbf{E-step}:
    \WHILE {convergence criterion not met }
        \STATE $l=1$.
        \STATE Choose a $\alpha_{k,i}^{(l-1)}$ (or equivalently choose a basis vector $[\hat{\boldsymbol{\Phi}}_k]_{:,i}$).
        \IF {$(q_{k,i}^{(l-1)})^2>p_{k,i}^{(l-1)}$ $\& $ $\alpha_{k,i}^{(l-1)}=\infty$}
            \STATE $\textbf{add}$ $\alpha_{k,i}^{(l)}$ to the model.
        \ELSIF {$(q_{k,i}^{(l-1)})^2>p_{k,i}^{(l-1)}$ $\& $ $\alpha_{k,i}^{(l-1)}\neq\infty$}
            \STATE $\textbf{re-estimate}$ $\alpha_{k,i}^{(l)}$.
        \ELSIF {$(q_{k,i}^{(l-1)})^2<p_{k,i}^{(l-1)}$}
                \STATE $\textbf{prune}$ $i$ from the model (set $\alpha_{k,i}^{(l)}=\infty$).
        \ENDIF
        \STATE Update $\boldsymbol{\mu}_k^{e,(l)}$ and $\boldsymbol{\Sigma}_k^{e,(l)}$.
        \STATE Update $p_{k,i}^{(l)}$ and $q_{k,i}^{(l)}$.
        \STATE $l\gets l+1$.
        \ENDWHILE
        \STATE Update $\mathbf{g}_k^{ul}$ by \eqref{eq:threeD2 Sigma}.
        \STATE \textbf{M-step}:
       \STATE Update $\boldsymbol{\beta}_k^{ul}$ by \eqref{eq:threeF beta}.
       \STATE Update $\boldsymbol{\upsilon}_k^{ul}$ by \eqref{eq:threeF nu}.
  \STATE  $l^{EM}\gets l^{EM}+1$.
  \ENDWHILE
  \RETURN $\mathbf{g}^{ul}_k$, $\boldsymbol{\beta}_k^{ul}$ and $\boldsymbol{\upsilon}_k^{ul}$.
 \end{algorithmic}
\end{algorithm}

\section{Effective Downlink Channel Reconstruction and Estimation over Delay-Doppler-Angle Domain }


\subsection{DL Channel Reconstruction at BS}

Within $[(n_1+(L_{cp}+N_t)(k-1))T_s,(n_1+(L_{cp}+N_t)k-1)T_s]$,
each user captures its UL channel parameter sets through achieving
$\{\hat{\tilde{\mathbf{G}}}_k^{ul}$, $\hat{\boldsymbol\beta}_k^{ul}$, $\hat{\boldsymbol\upsilon}_k^{ul}\}$. However, as shown in Fig. \ref{fig:Overall structure}, the OTFS-based transmission happens along DL within the interval
$[n_oT_s, n_oT_s+(L_D+L_{cp})N_D]$. Thus, we should utilize UL channel characteristics seen by BS within
the interval $[(n_1+(L_{cp}+N_t)(k-1))T_s,(n_1+(L_{cp}+N_t)k-1)T_s]$ to infer the time-varying
trajectory of the DL channels during the OTFS transmission. Thanks to the geometric propagation model in \eqref{eq:h_kln}
and \eqref{eq: h_ul_time}, and we can complete this task according to the following steps.


Firstly, we utilize the sparse channel gain matrix $\hat{\tilde{\mathbf{G}}}_k^{ul}$,
the Doppler shift vector $\hat{\boldsymbol\upsilon}_k^{ul}$
and the angle basis vector $\hat{\boldsymbol\beta}_k^{ul}$ to achieve the estimations of the $P$ UL parameter sets, i.e.,
$\{{\tau}_{k,p}^{ul}, \!{\nu}_{k,p}^{ul}, \!{\theta}_{k,p}^{ul},\!{h}_{k,p}^{ul}\}_{p=1}^P$.
Secondly, the obtained UL parameter sets are  utilized to construct the DL channel parameter sets, i.e.,
$\{{\tau}_{k,p},{\nu}_{k,p}, {\theta}_{k,p},{h}_{k,p}\}_{p=1}^P$,
which will be resorted to construct the delay-Doppler-angle domain channel $\bar{h}_{k,i,j,q}$ in the third step.

For the first step, as the channel gain matrix $\hat{\tilde{\mathbf{G}}}_k^{ul}$ is sparse, we can extract several non-zero points of $\hat{\tilde{\mathbf{G}}}_k^{ul}$ to represent the whole original channel gain. Intuitively, we can collect the points with most power successively, until the power efficiency reaches an acceptable rate \cite{MYLi_SBL_Time_varing}.
Then we can obtain the coordinate set for all the sparse points as
$\{(i^{no}_{k,p},j^{no}_{k,p}) \}_{p=1}^{P}$. Correspondingly, the UL channel parameters of the user $k$
can be derived from both $\{(i^{no}_{k,p},j^{no}_{k,p}) \}_{p=1}^{P}$ and $\{\hat{\tilde{\mathbf{G}}}_k^{ul}$, $\hat{\boldsymbol\beta}_k^{ul}$, $\hat{\boldsymbol\upsilon}_k^{ul}\}$ as
$\theta_{k,p}^{ul} = \vartheta_{i^{no}_{k,p}} + [\hat{\boldsymbol\beta}_k^{ul}]_{i^{no}_{k,p}}$,
$\nu_{k,p}^{ul} = [\hat{\boldsymbol\upsilon}_k^{ul}]_{j^{no}_{k,p}}$,
$\tau_{k,p}^{ul} = {j^{no}_{k,p}} T_s$, and
$h_{k,p}^{ul} e^{\jmath 2\pi \nu_{k,p}^{ul} (n_1+(L_{cp}+N_t)(k-1)+L_{cp} )T_s} = [\hat{\tilde{\mathbf{G}}}_k^{ul}]_{i^{no}_{k,p},j^{no}_{k,p}}$, $p = 1,2,\ldots, P$. Notice that the terms $h_{k,p}^{ul}$ and $e^{\jmath 2\pi \nu_{k,p}^{ul}(n_1+(L_{cp}+N_t)(k-1)+L_{cp} )T_s}$
can not be decoupled.

After the acquiring of the UL parameter sets $\{\!\tau_{k,p}^{ul}, \!{\nu}_{k,p}^{ul},\! {\theta}_{k,p}^{ul},\!{h}_{k,p}^{ul} e^{\jmath 2\pi \nu_{k,p}^{ul}
(n_1\!+\!(L_{cp}\!+\!N_t)(k\!-\!1)\!+\!L_{cp} )T_s} \!\}_{p=1}^P$,  we can derive the DL parameter sets
with the aid of the UL ones.
In the following, we will depict the derivation for the TDD/FDD modes, respectively.

\subsubsection{Deriving the Parameters in TDD Mode}
In the TDD mode, as there is the reciprocity between the UL/DL channels,
DL channel model parameters are the same of the UL ones.
So we can obtain that $\tau_{k,p}={j^{no}_{k,p}} T_s$, $\theta_{k,p} = \vartheta_{i^{no}_{k,p}} + [\hat{\boldsymbol\beta}_k^{ul}]_{i^{no}_{k,p}}$, $\nu_{k,p} = [\hat{\boldsymbol\upsilon}_k^{ul}]_{j^{no}_{k,p}}$.

\subsubsection{Deriving the Parameters in FDD Mode}

In the FDD mode, there is no reciprocity between UL/DL channels.
So DL channel parameters $\{\tau_{k,p}, \nu_{k,p}, \theta_{k,p}, h_{k,p}\}_{p=1}^P$  are not the same with the UL ones.
Fortunately, since the propagation paths of the radiowaves are reciprocal, we can derive $\{\tau_{k,p}, \nu_{k,p}, \theta_{k,p} \}_{p=1}^P $ with the aid of the UL ones as $\tau_{k,p}={j^{no}_{k,p}} T_s$, $\theta_{k,p} = \vartheta_{i^{no}_{k,p}} + [\hat{\boldsymbol\beta}_k^{ul}]_{i^{no}_{k,p}}$, and $\nu_{k,p} = \lambda^{ul} [\hat{\boldsymbol\upsilon}_k^{ul}]_{j^{no}_{k,p}} / \lambda$. But, the channel gains can not be exactly inferred.
Nevertheless, as presented in \eqref{eq:fiveB H^DDA}, the sparse property  of $\bar{h}_{k,i,j,q}$ is determined by $\{\tau_{k,p}, \nu_{k,p}, \theta_{k,p} \}_{p=1}^P$. Thus, we can decide the locations of the  dominant elements among all $\bar{h}_{k,i,j,q}$ with the UL channel parameters, but not their exact values.
However, the users can estimate $\bar{h}_{k,i,j,q}$ with low overhead and can feed back
these known CSI to help the BS calibrate the $P$ DL channel parameter sets.

Before illustrating the operations in the third step, we give the following observations about
$\bar{h}_{k,i,j,q}$. From \eqref{eq:fiveB H^DDS} and \eqref{eq:fiveB H^DDA}, it can be concluded
that $\bar{h}_{k,i,j,q}$ within the OTFS transmission
is exactly determined by the $\{\tau_{k,p}, \nu_{k,p}, \theta_{k,p}, h_{k,p}e^{\jmath 2\pi \nu_{k,p}n_o
T_s}\}_{p=1}^P$. So after the recovering of the parameter sets
$\{\tau_{k,p}, \nu_{k,p}, \theta_{k,p}, h_{k,p}e^{\jmath 2\pi \nu_{k,p}^{ul}(n_1+(L_{cp}+N_t)(k-1)+L_{cp} )T_s}\}_{p=1}^P$, we can only utilize  the phase rotation operation {$e^{\jmath 2\pi \nu_{k,p}^{ul}
(\frac{\lambda^{ul}}{\lambda}n_o-(n_1+(L_{cp}+N_t)(k-1)+L_{cp} ))T_s}$ to modify $h_{k,p}e^{\jmath 2\pi \nu_{k,p}^{ul} (n_1+(L_{cp}+N_t)(k-1)+L_{cp} )T_s}$}.

\subsection{DL Channel Estimation}

From \eqref{eq:fiveB H^DDA},
it can be checked that $\bar{h}_{k,i,j,q}$ is dominant with the index set
$\mathcal Q_k=\left\{(i_{k,p},j_{k,p},q_{k,p})\right\}_{p=1}^P$, where
\begin{align}
q_{k,p}=\lfloor M \frac{d \sin \theta_{k,p}}{\lambda}\rfloor,\quad
i_{k,p}=\lfloor\tau_{k,p} L_D\triangle f\rfloor,\quad 
j_{k,p}=\lfloor\nu_{k,p}N_DT\rfloor.\label{eq:signature}
\end{align}

Obviously, $(i_{k,p},j_{k,p},q_{k,p})$ corresponds to the path $(\tau_{k,p},\nu_{k,p},\theta_{k,p})$ and can be treated as this path's delay-Doppler-angle signature.

\begin{figure}[!htp]
	\centering
    \subfigure[Pilot at BS]{
    \begin{minipage}{4cm}
    \centering
	\includegraphics[width=4cm]{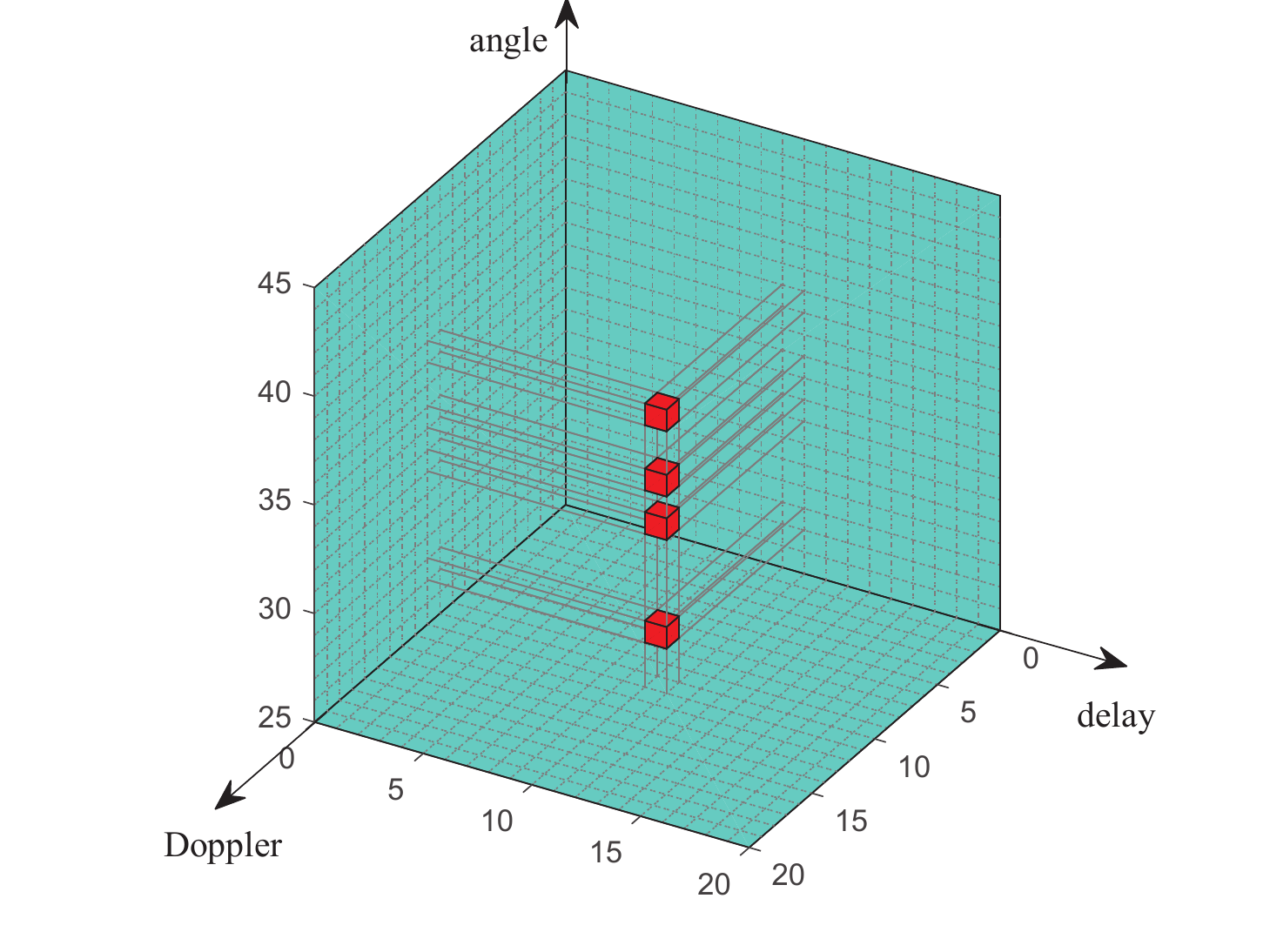}
    \end{minipage}
    }
    \subfigure[3D channel]{
    \begin{minipage}{4cm}
    \centering
	\includegraphics[width=4cm]{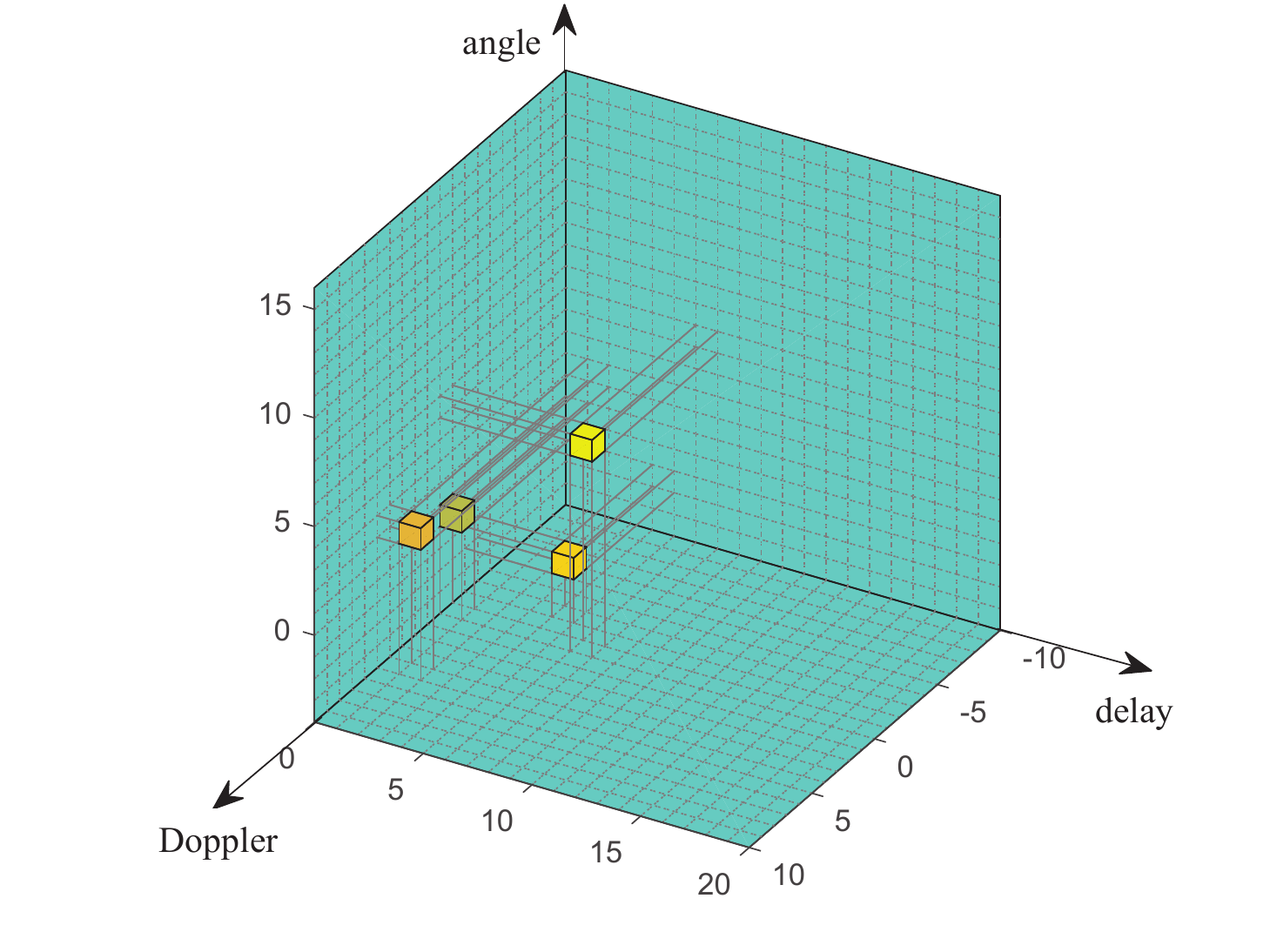}
    \end{minipage}
    }
    \subfigure[Observation at user side]{
    \begin{minipage}{4cm}
    \centering
	\includegraphics[width=4cm]{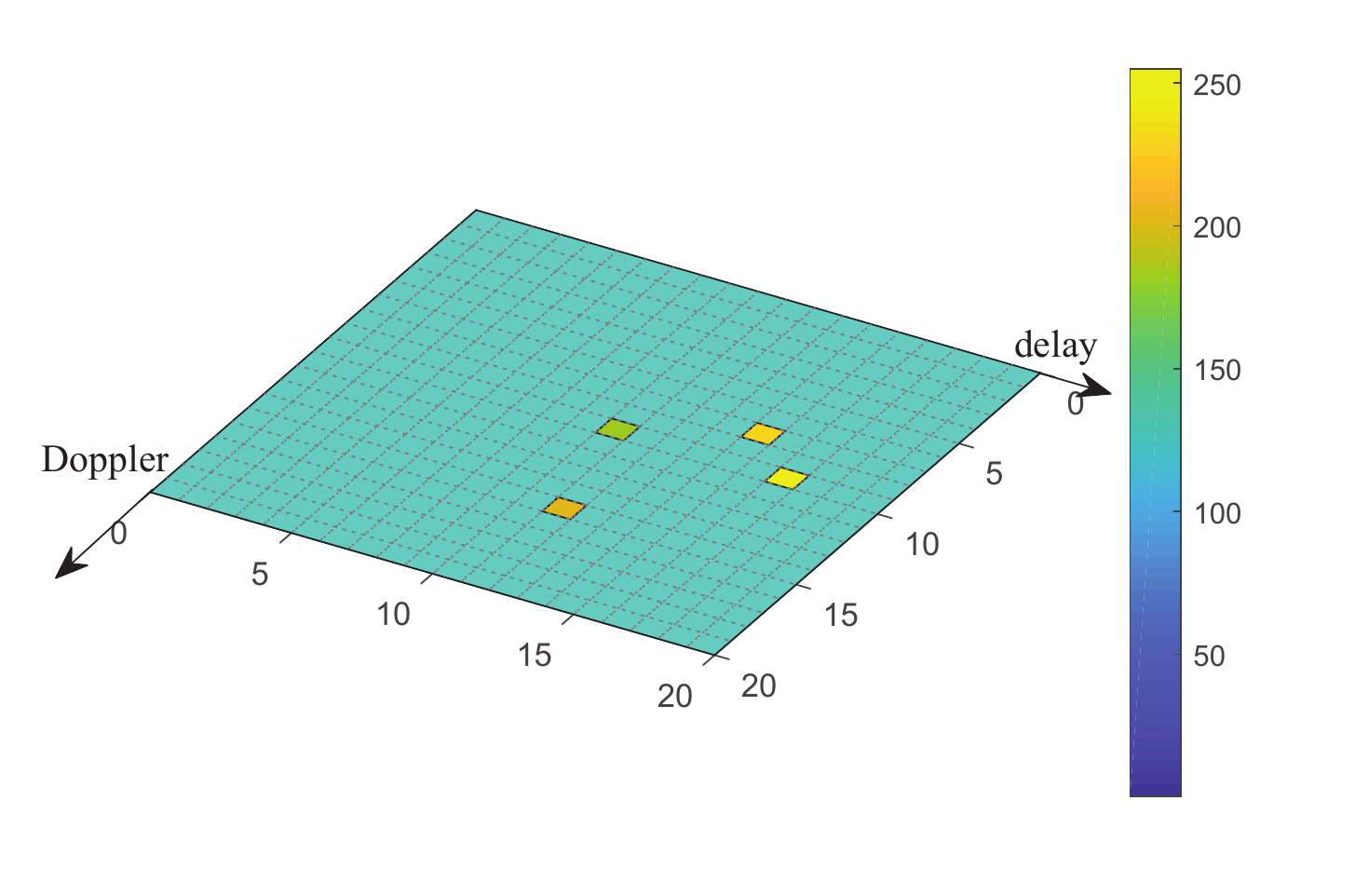}
    \end{minipage}
    }
	\caption{The channel dispersion over the delay-Doppler-angle domain.}
	\label{fig:channel_dispense}
\end{figure}

{From \eqref{eq:fiveA Y^DD}, we can obtain the following observation:
If the BS sends one effective symbol at $x_{l,n,m}$, the user $k$ can only receive the information of this symbol at $y_{k,l',n'}$ with the index set $(l'=l+i_{k,p},n'=n+j_{k,p})$, where $(i_{k,p},j_{k,p},q_{k,p})\in\mathcal Q_k$.
In other words, the OTFS scheme possesses the energy dispersion within
the delay-Doppler domain. As shown in Fig. \ref{fig:channel_dispense}, once we can achieve the delay and the Doppler frequencies,
the exact dispersion locations within the OTFS block can be determined, which may help us decrease
 the observation dimension.}

{Let us define the $M\!\!\times \!\!1$ vectors $\bar{\mathbf{h}}_{k,l,n}\!\!\!=\!\!\![\!\bar{h}_{k,l,n,\!-\!\frac{M}{2}}\!,\!\ldots\!,\!\bar{h}_{k,l,n,0}\!,\!\ldots\!,\!\bar{h}_{k,l,n,\frac{M}{2}\!-\!1}\!]^T$,
$\mathbf x_{l,n}\!\!\!=\!\!\![\!x_{l,n,0},\! \ldots,\!x_{l,n,M\!-\!1}\!]^T$, $\bar{\mathbf x}_{l,n}=[\bar{x}_{l,n,0},\ldots,\bar{x}_{l,n,M-1}]^T= \mathbf F_{M}^*\mathbf x_{l,n}$, and $P\times 1$ vector
$\mathbf{\bar h}^{no}_k=[\bar h_{k,i_{k,1},j_{k,1},q_{k,1}},\ldots,\bar h_{k,i_{k,P},j_{k,P},q_{k,P}}]^T$ for further use,
where $l\in[0,L_D-1]$ and $n\in[0,N_D-1]$.
Correspondingly, we give the $L_D\times N_D$ matrices $\mathbf{\bar X}_s$, whose
$(l,n)$-th element is $[\bar{\mathbf x}_{l,n}]_s$ and $s\in[0,M-1]$.
Moreover, the angle signature set for the user $k$ is given as $\mathcal Q_k^a=\!\{q_{k,p}\}_{p=1}^P$,
and the maximum dispersion length along the delay and Doppler directions are separately written as}
\begin{align}
l^G=\max\{\{i_{k,p}\}_{p=1}^P\}_{k=1}^K,\kern 10pt n^G=\max\{\{|j_{k,p}|\}_{p=1}^P\}_{k=1}^K.
\end{align}
Then, we can rewrite \eqref{eq:fiveA Y^DD} as follows
\begin{align}
{y_{k,l,n}}
{=}&{\sum_{m=0}^{M-1}\sum_{p=1}^P e^{\jmath2\pi\frac{(l-i_{k,p})j_{k,p}}{N_D(L_D+L_{cp})}}\tilde h_{k,i_{k,p},j_{k,p},m} x_{(l-i_{k,p})_{L_D},\langle n-N_D/2-j_{k,p}\rangle+N_D/2,m}+w_{k,l,n}}\notag\\
{=}&{\sum_{p=1}^P \!e^{\jmath2\pi\frac{(l-i_{k,p})j_{k,p}}{N_D(L_D+L_{cp})}}\left(\mathbf F^H_M\mathbf{\bar h}_{k,i_{k,p},j_{k,p}}\right)^T\left( \mathbf{F}_M^T\bar{\mathbf{x}}_{(l-i_{k,p})_{L_D},\langle n-N_D/2-j_{k,p}\rangle+N_D/2}\right)+w_{k,l,n}}\notag\\
{=}&{\sum_{p=1}^P e^{\jmath2\pi\frac{(l-i_{k,p})j_{k,p}}{N_D(L_D+L_{cp})}}\!{\bar{x}_{(l-i_{k,p})_{L_D},\langle n-N_D/2-j_{k,p}\rangle+N_D/2,q_{k,p}+M/2}}\bar{h}_{k,i_{k,p},j_{k,p},q_{k,p}}+w_{k,l,n}}.\label{eq:DLy2}
\end{align}
{According to the characteristics of 3D channels $\bar{h}_{k,i,j,q}$, we design three training schemes
and optimize the pilot pattern over the 3D cubic space $\{(l,n,s)|l\in[0,L_D-1], n\in[0,N_D-1], s\in[0,M-1]\}$,
which contains $M$ layers along the angle direction. Explicitly, the $s$-th layer is $\mathbf{\bar X}_s$.}

\subsubsection{{The Case that All Paths of the User $k$ are Orthogonal over Delay-Doppler domain}}

{Interestingly, we analyze one special case, where
different scattering paths of the user $k$ have distinguished delays or Doppler frequencies.
To fully exploit the energy dispersion effective of the OTFS, we adopt the embedded pilot structure \cite{OTFSesti1} to
estimate $\bar{h}_{k,i,j,q}$ and place
only one effective pilot within fixed rectangle area of $\mathbf {\bar X}_s$.
Without loss of generality, this training part is fixed in the region $\{(l,n)|l\in[0,\ldots, l^G],$ $n\in[0,\ldots,2n^G]\}$ of {$\mathbf{\bar X}_s$}. With the previous results, it can be checked
that the user $k$ can receive signal from $P$ layers within the 3D cubic region, i.e.,
$\bar{\mathbf X}_{q_{k,p}+M/2}$, $p\!=\!1,2,\ldots, P$. Taking all
the $K$ users into consideration, we would fix only one non-zero pilot
at ${\bar x}_{0,n^G,q+M/2}\!\!=\!\!\sqrt{{\sigma_p^2}}$, where $q\!\!\in\!\!\{\mathcal Q^a_1\!\cup\!\mathcal Q^a_2\!\cup\!\ldots\!\cup\!\mathcal Q^a_K\}$ and $\sigma_p^2$ is the training power.}

%



{With (\ref{eq:DLy2}), it can be determined that the user $k$ achieves the non-zero training power
at $P$ grids $\{y_{k,i_{k,1},n^G+j_{k,1}},y_{k,i_{k,2},n^G+j_{k,2}},\ldots,y_{k,i_{k,P},n^G+j_{k,P}}\}$, which separately correspond to the elements in $\mathbf{\bar h}^{no}_k$.}
{Moreover, $y_{k,i_{k,p},n^G_k+j_{k,p}}$, $p\in[1,P]$, can be given as}
\begin{align}
{y_{k,i_{k,p},n^G_k+j_{k,p}}
=}&{\bar{x}_{0,n_k^G,q_{k,p}+M/2}} {\bar{h}_{k,i_{k,p},j_{k,p},q_{k,p}}+w_{k,i_{k,p},n^G_k+j_{k,p}},}\notag\\
{=}&{\sqrt{{\sigma_p^2}}\bar{h}_{k,i_{k,p},j_{k,p},q_{k,p}}+w_{k,i_{k,p},n^G_k+j_{k,p}}.}
\end{align}
{Then, we can obtain}
\begin{align}
{\widehat{\bar{h}}_{k,i_{k,p},j_{k,p},q_{k,p}}=\sqrt{\frac{1}{\sigma_p^2}}y_{k,i_{k,p},n^G_k+j_{k,p}}.
\label{eq:est_channel_dDa}}
\end{align}

\subsubsection{{The Case that Different Paths of the User $k$ are Orthogonal over Angle Domain}}

We consider the case where different scattering paths of the user $k$  have distinguished angles, $k\!\!=\!\!1,2,\ldots,K$.
For clear explanation, we firstly consider one user $k$.
Without loss of generality, we assume that BS would send the user $k$ effective data from
the rectangle region of the $P$ effective layers, i.e.,
$\bar{\mathbf X}_{q_{k,p}\!+\!\frac{M}{2}}$, $p\!\!=\!\!1,2,\ldots, P$.
Specially, in $\bar{\mathbf X}_{q_{k,p}\!+\!\frac{M}{2}}$, the effective data lie in the
region
$\{(l,n)|l\!\!\in\!\![l_{k,p},l_{k,p}\!+\!W_d\!-\!1],n\!\!\in\!\![n_{k,p},n_{k,p}\!+\!W_D\!-\!1]\}$,
where $W_d$ and $W_D$ are the maximum lengthes of this region along the delay and Doppler directions, respectively.
Then, the user $k$ would receive $P$ signal components along different scattering paths, and each component has
the same size with $\mathbf Y_k$. Here, we denote the $p$-th received signal part from $\bar{\mathbf X}_{q_{k,p}\!+\!\frac{M}{2}}$ as $\mathbf{Y}_{k,p}$, which experiences the channel $\bar{h}_{k,i_{k,p},j_{k,p},q_{k,p}}$.
Correspondingly, $[\mathbf Y_{k,p}]_{l,n}$ can be derived from \eqref{eq:DLy2} as
\begin{small}\begin{align}
[\mathbf Y_{k,p}]_{l,n}
=e^{\jmath2\pi\frac{(l-i_{k,p})j_{k,p}}{N_D(L_D+L_{cp})}}\bar{h}_{k,i_{k,p},j_{k,p},q_{k,p}}
[\bar{\mathbf{X}}_{q_{k,p}+M/2}]_{(l-i_{k,p})_{L_D},\langle n-N_D/2-j_{k,p}\rangle +N_D/2},\label{eq:Y_point}
\end{align}\end{small}{which means that each grid in $\mathbf Y_{k,p}$ is associated only one symbol in $\mathbf{\overline X}_{q_{k,p}+M/2}$. Then,
{$\mathbf Y_k$} can be rewritten as}
\begin{align}
\mathbf Y_k=\sum_{p=1}^P \mathbf Y_{k,p}+\mathbf W_k,\label{eq:Y_area}
\end{align}
{where the $(l,n)$-th element of $\mathbf{W}_k$ is equal to $w_{k,l,n}$.}

\begin{figure}[!t]
	\centering
	\includegraphics[width=4in]{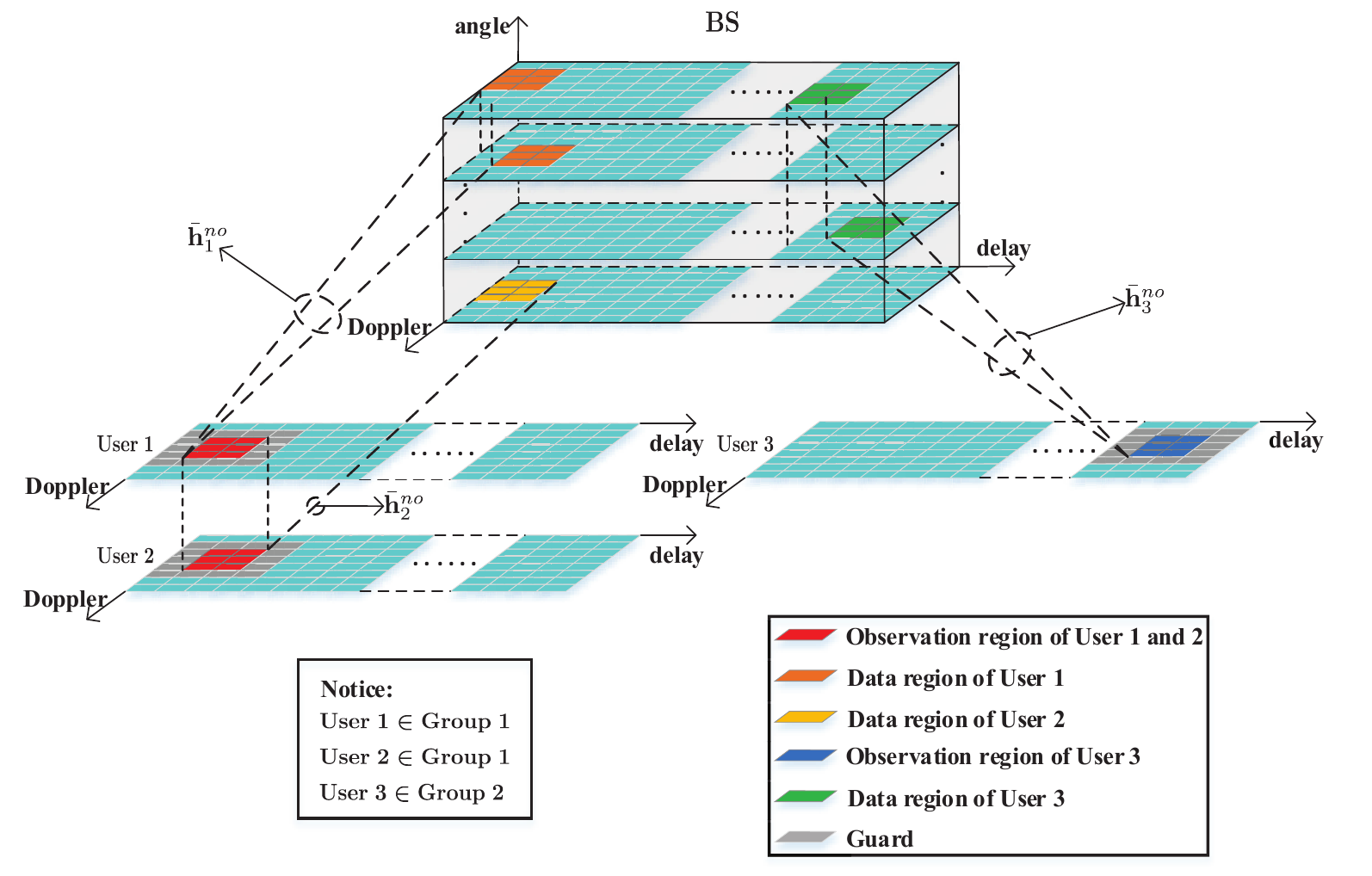}
	\caption{{Path scheduling under the multi-user case.}}
	\label{fig:pre-compensation}
\end{figure}

{To simply capture the path diversity,
we align the associated grids for
\begin{small}$[\mathbf{\bar X}_{q_{k,1}+M/2}]_{l_{k,1}+u,n_{k,1}+v}$,
$[\mathbf{\bar X}_{q_{k,2}+M/2}]_{l_{k,2}+u,n_{k,2}+v}$$\ldots$
$[\mathbf{\bar X}_{q_{k,P}+M/2}]_{l_{k,P}+u,n_{k,P}+v}$\end{small}
to the same
position, $u\in[0,W_d-1], v\in[0,W_D-1]$. Hence, the related positions for
$P$ effective data rectangles over the 3D cubic area should satisfy}
{$(l_{k,1}+i_{k,1})_{L_D}=(l_{k,2}+i_{k,2})_{L_D}=\cdots=(l_{k,P}+i_{k,P})_{L_D}$ and $(n_{k,1}+j_{k,1})_{N_D}=(n_{k,2}+j_{k,2})_{N_D}=\cdots=(n_{k,P}+j_{k,P})_{N_D}$, which is shown in Fig. \ref{fig:pre-compensation}.}

Without loss of generality, we restrict the effective observation region  of user $k$ in one rectangle of $\mathbf Y_k$, which
can be described by the grid set $\mathcal A_k^r$ as
\begin{align}\label{trans_mtx}
{\mathcal A_{k}^r = \left\{(l,n) | l \in [l_k, l_k+W_d-1], n \in [n_k, n_k+W_D-1] \right\}},
\end{align}
{where the element in $\mathcal A_{k}^r$ is the index of effective observation grid.
Then, for $(l,n)\in \mathcal A_k^r$, it receives the information from $P$ grids within the 3D cubic area, whose
specific locations in $\bar{\mathbf X}_s$ can be written as $\{((l-i_{k,1})_{L_D}, \langle n-N_D/2-j_{k,1}\rangle+N_D/2, q_{k,1}+ M/2), ((l-i_{k,2})_{L_D}, \langle n-N_D/2-j_{k,2}\rangle+N_D/2, q_{k,2}+ M/2), \ldots, ((l-i_{k,P})_{L_D}, \langle n-N_D/2-j_{k,P} \rangle+N_D/2, q_{k,P}+ M/2)\}$. Then, with respect to $\mathcal A_k^r$, the effective transmission region of the $\mathbf{\bar X}_{q_{k.p}+\frac{M}{2}}$ can be written as}
%
%
\begin{align}
{\mathcal{A}^t_{k,p}=}\Big\{&(l,n)|l=(l'-i_{k,p})_{L_D},n=\langle n'-N_D/2-j_{k,p}\rangle+N_D/2,(l',n')\in\mathcal{A}^r_k\Big\}.
\end{align}
{Thus, combine \eqref{eq:Y_point} and \eqref{eq:Y_area}, we obtain}
\begin{small}\begin{align}
{[\mathbf{Y}_{k}]_{\mathcal{A}^r_k}}
=&\sum_{p=1}^P\underbrace{\bar{\boldsymbol{\Phi}}_{q_{k,p}+M/2} \odot [\bar{\mathbf{X}}_{q_{k,p}+M/2}]_{\mathcal{A}^t_{k,p}}}_{\bar{\mathbf{X}}^e_{k,q_{k,p}+M/2}}\bar{h}_{k,i_{k,p},j_{k,p},q_{k,p}}+[\mathbf{W}_{k}]_{\mathcal{A}^r_k}\notag\\
=&[\bar{\mathbf{X}}^e_{k,q_{k,1}+M/2},\ldots,\bar{\mathbf{X}}^e_{k,q_{k,P}+M/2}]\bar{\mathbf{h}}_{k}^{no}+[\mathbf{W}_{k}]_{\mathcal{A}^r_k},\label{eq:Y_rectangle}
\end{align}\end{small}{where}
{$[\bar{\boldsymbol{\Phi}}_{q_{k,p}\!+\!M/2}]_{u,v}\!=\!e^{\jmath2\pi\frac{(l_k\!+\!u\!-\!i_{k,p})j_{k,p}}{N_D(L_D\!+\!L_{cp})}}$, $u\!\in\![0,W_d\!-\!1]$, $v\in[0,W_D\!-\!1]$ and the $W_d\!\times \!W_D$ matrix $\bar{\mathbf{X}}^e_{k,q_{k,p}+M/2}$ is given above}.

{Since only $P$ unknown channel gains should be estimated,
we choose $P$ symbols in $\mathcal A_{k}^r$ as the training observation grids, whose
indexes can be listed as $\{(l_k,n_k),(l_k,n_k+1),\ldots, (l_k,n_k+P-1)\}$. Correspondingly, we have}
\begin{align}
{[\mathbf{Y}_{k}]_{l_k,n_k:n_k+P-1}}
=\underbrace{\big[[\bar{\mathbf{X}}^e_{k,q_{k,1}+M/2}]_{0,0:P-1},\ldots,[\bar{\mathbf{X}}^e_{k,q_{k,P}+M/2}]_{0,0:P-1}\big]}_{\mathbf{T}_k}\bar{\mathbf{h}}_{k}^{no}+[\mathbf{W}_{k}]_{l_k,n_k:n_k+P-1},
\end{align}
{where the $P\times P$ training matrix $\mathbf T_k$ is defined in the above equation.
Then, with LS estimator,
we can obtain}
\begin{align}
{\bar{\mathbf{h}}_{k}^{no}=\mathbf{T}_k^{-1}[\mathbf{Y}_{k}]_{l_k,n_k:n_k+P-1},}
\end{align}
{where the optimal training matrix should satisfy $\mathbf T_k^H\mathbf T_k=\sigma_p^2\mathbf I_P$.}


{The above strategy can be extended for the multi-user case.
Due to the inter-user interference, it is necessary to schedule $\mathcal{A}^t_{k,p}$
to assure the effective transmitting regions for different users do not overlap in the 3D cubic area.
To fully exploit the super resolution over the
angle domain, we schedule the users with two steps: within the first step, the users would be separated
over the angle domain, while in the second step, we further separate the users, that have overlapped
angle signatures, in the delay-Doppler domain.
Then, the users with non-overlapped angle signatures sets are allocated to the same group $\mathcal G_g$ $(g=1,2,\ldots,G)$, i.e.,}
\begin{align}
{\mathcal Q_{k_1}^a \cap \mathcal Q_{k_2}^a = \emptyset,
\kern 10pt
\text{dist}(\mathcal Q_{k_1}^a, \mathcal Q_{k_2}^a) \ge D_{\theta},}
\end{align}
{where $\text{dist}(\mathcal{Q}^a_{k_1}\!, \mathcal{Q}^a_{k_2}) \!\!\triangleq\!\! \min\! |q_{k_1,p}\!-\!q_{k_2,p}|, \forall\! q_{k_1,p}\! \in \!\mathcal{Q}^a_{k_1}, \forall\! q_{k_2,p} \!\in\! \mathcal{Q}^a_{k_2}$, and the protection gap $D_{\theta}$ is added to mitigate the effect of the channel power leakage along
the angle direction. For $k_1, k_2\in \mathcal G_g$, we assign them the same observation region over the delay-Doppler domain, i.e.,
$\mathcal{A}^r_{k_1}=\mathcal{A}^r_{k_2}=\mathcal{A}^r_{\mathcal G_g}$, but different angle grids, i.e., $\mathcal Q_{k_1}^a\cap\mathcal Q_{k_2}^a=\emptyset$, $k_1\neq k_2$.}

\begin{figure}[!htp]
	\centering
	\includegraphics[width=3.5in]{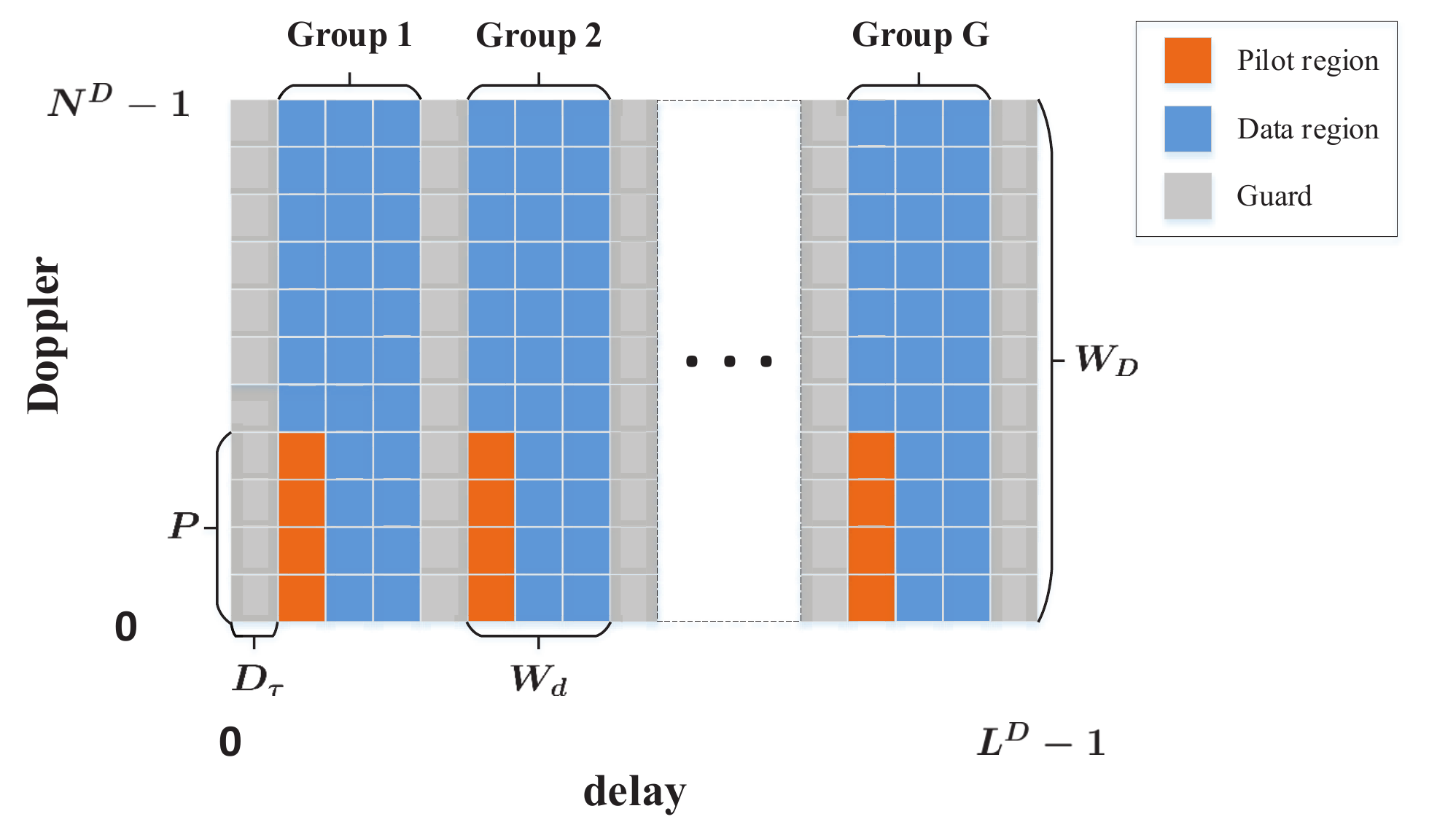}
	\caption{{The observation regions for different users over the delay-Doppler domain.}}
	\label{fig:case2_pilot_design}
\end{figure}

{With different user groups $\mathcal G_{g_1}$, $\mathcal G_{g_2}$, we assign distinguished delay-Doppler domain observation regions  to
satisfy the following constraints:}
\begin{align}
{\mathcal A^r_{\mathcal G_{g_1}}\cap\mathcal A^r_{\mathcal G_{g_2}}=\emptyset,
\kern 10pt
\text{dist}(\mathcal A^r_{\mathcal G_{g_1}}, \mathcal A^r_{\mathcal G_{g_2}}) \succ \{D_\tau, D_\nu\},}
\end{align}
{where $\text{dist}(\mathcal{A}^r_{g_1},\mathcal{A}^r_{g_2})\succ\{D_\tau, D_\nu\}$ means $\min |l_1\!-\!l_2|\!\geq\! D_\tau$ or $\min\! |n_1\!-\!n_2|\!\geq \!D_\nu$, $\forall (l_1,n_1) \!\in\! \mathcal{A}^r_{g_1}, \forall (l_2,n_2) \!\in\! \mathcal{A}^r_{g_2}$,
and the guard gaps $D_{\tau}$ and $D_{\nu}$ are mainly utilized to combat the dispersion effect of
the 3D channels $\bar h_{k,i,j,q}$ along the delay and Doppler directions, respectively.}
{Typically, we can separately set $D_{\tau}$ and $D_{\nu}$ as
$D_\tau=l^G$ and $D_\nu=2n^G$. One typical example is presented in Fig. \ref{fig:case2_pilot_design}. After scheduling, the BS can
map different users' respective data with the scheduled delay-Doppler-angle domain grids,
send the different data to the users within
the same OTFS block, and occupy different 3D resources. Then, the users can parallel
demap and decode respective data without inter-user interference.}


\subsubsection{{General Case}}

{Unfortunately, in practice, all paths of one user
may be not distinguished over the angle or delay-Doppler domains.
Theoretically, we can schedule the delay-Doppler-angle domain resources according
to the overlapping situation of the paths and achieve effective channel estimation schemes, which is beyond
the scope of this paper. Nonetheless, we present one feasible channel recovering method for this general case.}

%
{Taking all the $K$ users into consideration, we would fix the effective pilots of $\bar{\mathbf X}_{q\!+\!M/2}$ in the region $\{(l,n) | l \!\in \!(l_s, l_s\!+\!H_d\!-\!1), n\!\in\! (n_s\!+\!H_D\!-\!1)\}$,
where $q\!\in\!\{\mathcal Q^a_1\!\cup\!\mathcal Q^a_2\!\cup\!\ldots\!\cup\!\mathcal Q^a_K\}$, and $l_s$, $n_s$ separately denote the left and the bottom bounds of the effective pilot region within $\bar{\mathbf X}_s$. Moreover, $H_d$, $H_D$ are the maximum widths of the pilot region along the delay and Doppler axes, respectively.
Moreover, we assume that the effective pilot region satisfies $l\!+\!i_{k,p}\!<\!L_D\!-\!1$ and $0\!<\!n\!+\!j_{k,p}\!<\!N_D\!-\!1$, where
$\{{(l,n)} | l\! \!\in\!\! (l_s, l_s\!+\!H_d\!-\!1), n\!\in \!(n_s\!+\!H_D\!-\!1)\}$, $k\!\in\![1,K]$, and $p\!\!\in\!\![1,P]$.
Under this case, the observation regions for the pilots at  all users are continuous regions.}

{With \eqref{eq:DLy2} and the delay-Doppler-angle signature for the user $k$, we can check that
$y_{k,l,n}$ may have the non-zero entries within the region $\{(l,n)|
l\in[l_s+i_k^{min},l_s+H_d-1+l^G],n\in[n_s+j_k^{min},n_s+H_D-1+j_k^{max}]\}$, where
$i^{min}_k=\min~\{i_{k,p}\}_{p=1}^P$, $j_k^{min}=\min~\{j_{k,p}\}_{p=1}^P$, and
$j_k^{max}=\max~\{j_{k,p}\}_{p=1}^P$.
For simplicity, we rewrite \eqref{eq:DLy2} into the vector-matrix form. For user $k$, we can arrange $y_{k,l,n}$
within the region  $\{(l,n)|
l\in[l_s+i_k^{min},l_s+H_d-1+l^G],n\in[n_s+j_k^{min},n_s+H_D-1+j_k^{max}]\}$ into one column vector
$\mathbf{y}^{e}_k\in\mathcal{C}^{(H_d+l^G-i_k^{min})(H_D+j_k^{max}-j_k^{min})\times 1}$, whose $\left((\!H_D\!+\!j_k^{max}\!-\!j_k^{min}\!)\!(\!l\!-\!(\!l_s\!+\!i_k^{min}\!)\!)\!+\!n\!-\!(\!n_s\!+\!j_k^{min}\!)\right)$-th elements equal to $y_{k,l,n}$. Finally, we can obtain the rewritten received signal \eqref{eq:DLy2} as}
\begin{align}
{\mathbf{y}^{e}_k=\underbrace{\mathbf{V}^e_k\odot\bar{\mathbf{X}}^e_k}_{\boldsymbol{\Psi}^e_k}\mathbf{\bar h}^{no}_k+\mathbf{w}^e_k,}\label{eq:sixA1 y_c1}
\end{align}
{where the $\bar{\mathbf{X}}^e_k\in\mathcal{C}^{(H_d+l^G-i_k^{min})(H_D+j_k^{max}-j_k^{min})\times P}$ is the two-dimensional periodic convolution matrix with the
$\left((H_D+j_k^{max}-j_k^{min})(l-(l_s+i_k^{min}))+n-(n_s+j_k^{min}),p-1\right)$-th element of $\bar{\mathbf{X}}^e_k$ being equal to $\bar{x}_{l-i_{k,p},n-j_{k,p},q_{k,p}}$, $\mathbf{V}^e_k\in\mathcal{C}^{(H_d+l^G-i_k^{min})(H_D+j_k^{max}-j_k^{min})\times P}$ is a matrix with the $\left((H_D+\!j_k^{max}-j_k^{min})(l-(l_s+i_k^{min}))+n-(n_s+j_k^{min}),p-1\right)$-th element being $e^{\jmath2\pi\frac{(l-i_{k,p})j_{k,p}}{N_D(L_D+L_{cp})}}$,
and $\mathbf{w}^e_k$ is the AWGN vector, whose element has zero mean and variance $\sigma^2$. Finally, by using LS method we can obtain}
\begin{align}
{\hat{\bar{\mathbf{h}}}_k^{no}=\left((\boldsymbol{\Psi}^e_k)^T\boldsymbol{\Psi}^e_k\right)^{-1}(\boldsymbol{\Psi}^e_k)^T\mathbf{y}^e_k,}\label{eq:h_no}
\end{align}
{where the matrix $(\boldsymbol{\Psi}^e_k)^T\boldsymbol{\Psi}^e_k$ is low-dimensional.}


\subsection{Pilot Overhead Analysis}


{In our proposed DL channel estimation scheme, the training overhead comes from two parts: the UL channel parameter extraction and the DL 3D channel recovering, where the former happens over the frequency-time-antenna domain while the latter lies in the delay-Doppler-angle domain.
With respect to the first part, we utilize the time domain training sequence of length $K(L_{cp}+N_t)$.
Then, along the DL, we construct three training schemes over delay-Doppler-angle domain according to the characteristics of the scattering paths.
When all paths of any user can be separated over the delay-Doppler domain, no more than $\min\{KP,M\}$ grids over the delay-Doppler-angle
domain are needed. Moreover, if all paths of any user can be distinguished over the delay-Doppler domain, it takes us $KP^2$  grids in
the 3D cubic resource region to estimate the gains for all DL scattering paths. With respect to the general case,
the pilot overhead is $\min\{KPH_dH_D,MH_dH_D\}$ at most, where $H_d$ and $H_D$ can be appropriately selected according to (\ref{eq:h_no}).
It is noticed that the work in \cite{OTFSesti2} only considers DL. However, in practice, the communication happens along both DL and UL.
If \cite{OTFSesti2} analyze the UL OTFS channel estimation, the similar pilot overhead with ours would be utilized. Moreover, the authors in \cite{OTFSesti2} only focus the channel estimation but not the data transmission process for multiple users. In our work, we have given some feasible schemes for serving  multiple users with the massive MIMO-OTFS. The proper path scheduling  algorithm is given to avoid the inter-user interference.}

\section{Simulation Results}
In this section, we evaluate the performance of our algorithm through numerical simulation.
Unless otherwise specified, we  consider the TDD model.
The number of BS's antennas is $M=64$, {the carrier frequency is $6$~GHz},
and the antenna spacing $d$ is set as the half wavelength. {The $K$ users move at the speed}
$v_s\in[200,400]$~Km/h, and their maximal Doppler shift frequency is $2.22$ kHz.
With respect to the massive MIMO-OTFS scheme, the number of the subcarriers is $L_D=512$, the number of the OFDM symbols within one OTFS block is $N_D=128$, the length of CP is $L_{cp}=32$,
and the sampling time period $T_s=\frac{1}{20\ \text{MHz}}$.
Correspondingly, the resolutions over the delay and the Doppler domains are $T_s$ and $\frac{1}{(L_D+L_{cp})N_DT_s}\approx287$ Hz, respectively.
For the high mobility channel parameters, $\tau_{k,p}$ is randomly chosen from
$\{0,T_s,2T_s,\ldots,15T_s\}$,  while $\theta_{k,p}$ and $\nu_{k,p}$ are
uniformly distributed within $[-10^\circ,50^\circ]$ and $[-2.22,2.22]$ kHz, respectively.
Moreover, the power of multi-path channel gain $h_{k,p}$ is normalized as 1, i.e.,
$\sum\limits_{p=1}^P\mathbb E\{|h_{k,p}|^2\}=1$.
Furthermore, in our simulations, the angle grids $\vartheta_n$ lie uniformly within $[-90^\circ,90^\circ]$, and the number of the angle grids is
$90$. For the delay grids, the total number is $L=20$, and the delay girds can be written as $\{0,T_s,\ldots, 19T_s\}$.


The signal-to-noise ratio (SNR) is expressed as $\text{SNR}=10\log_{10} \sigma_p^2/\sigma^2$ dB.
Here, we use the normalized mean square error (MSE) for both UL channel parameters
$\mathbf{g}_k^{ul}$, $\boldsymbol{\beta}_k^{ul}$, $\boldsymbol{\upsilon}_k^{ul}$, $\bar{\mathbf h}^{no}_k$ and
DL delay-Doppler-angle channel vector $\bar{\mathbf h}^{no}_k$ as the performance metric, which is defined as
\begin{align} \text{MSE}_{\mathbf x}=\mathbb E\left\{\frac{||\hat{\mathbf x}-\mathbf x||^2}{||\mathbf x||^2}\right\},\mathbf x=\mathbf{g}_k^{ul},\boldsymbol{\beta}_k^{ul},\boldsymbol{\upsilon}_k^{ul},\bar{\mathbf h}^{no}_k,
\end{align}
where $\hat{\mathbf{x}}$ is the estimation of $\mathbf x$.

%
%

\begin{figure}[!htp]
	\centering
	\includegraphics[width=3in]{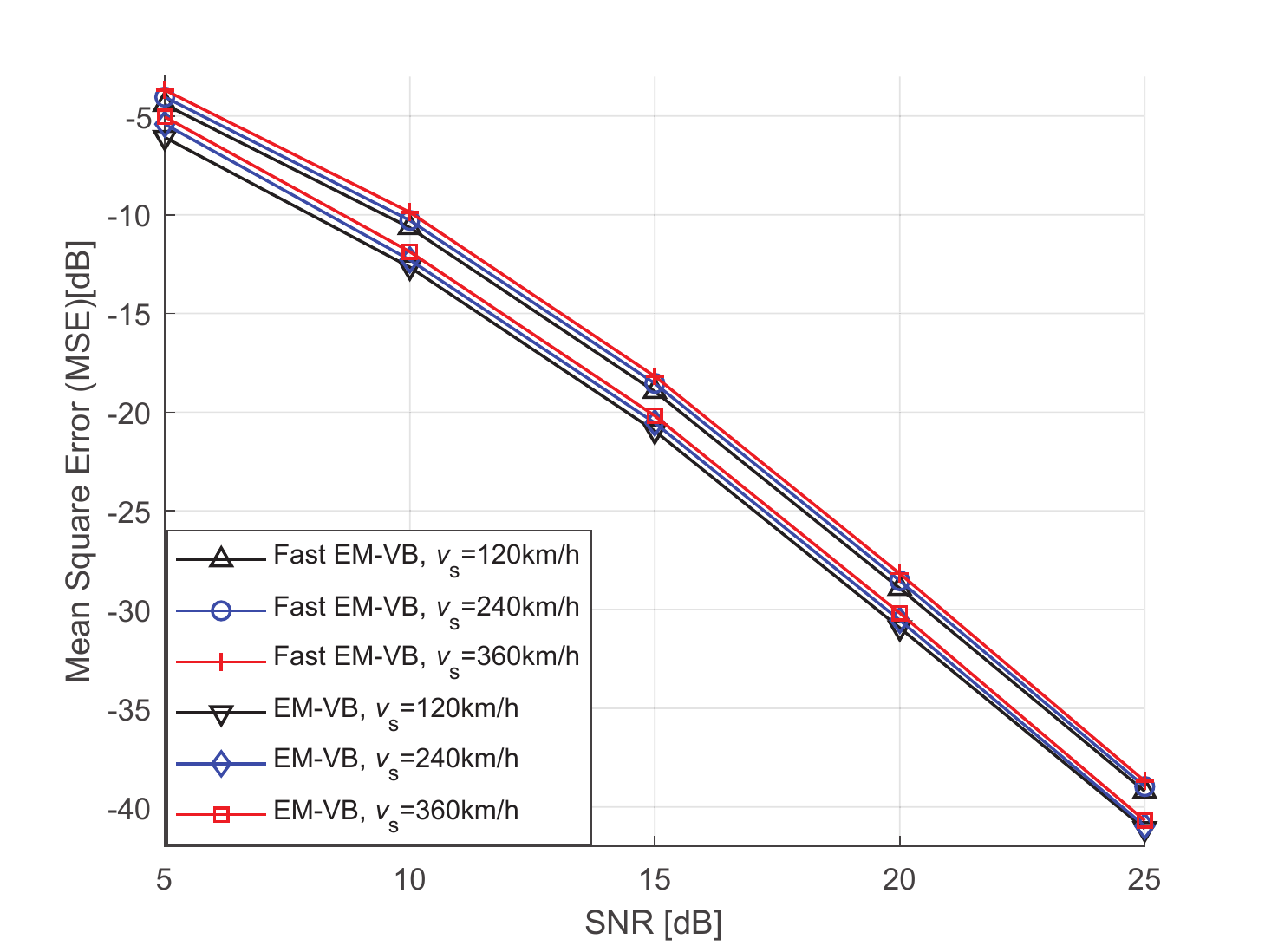}
	\caption{The MSEs of $\mathbf g_k^{ul}$ versus SNRs at different users' moving speeds.}
	\label{fig:ce_sweep_SNR_change_V}
\end{figure}

\begin{figure}[!htp]
	\centering
	\includegraphics[width=3in]{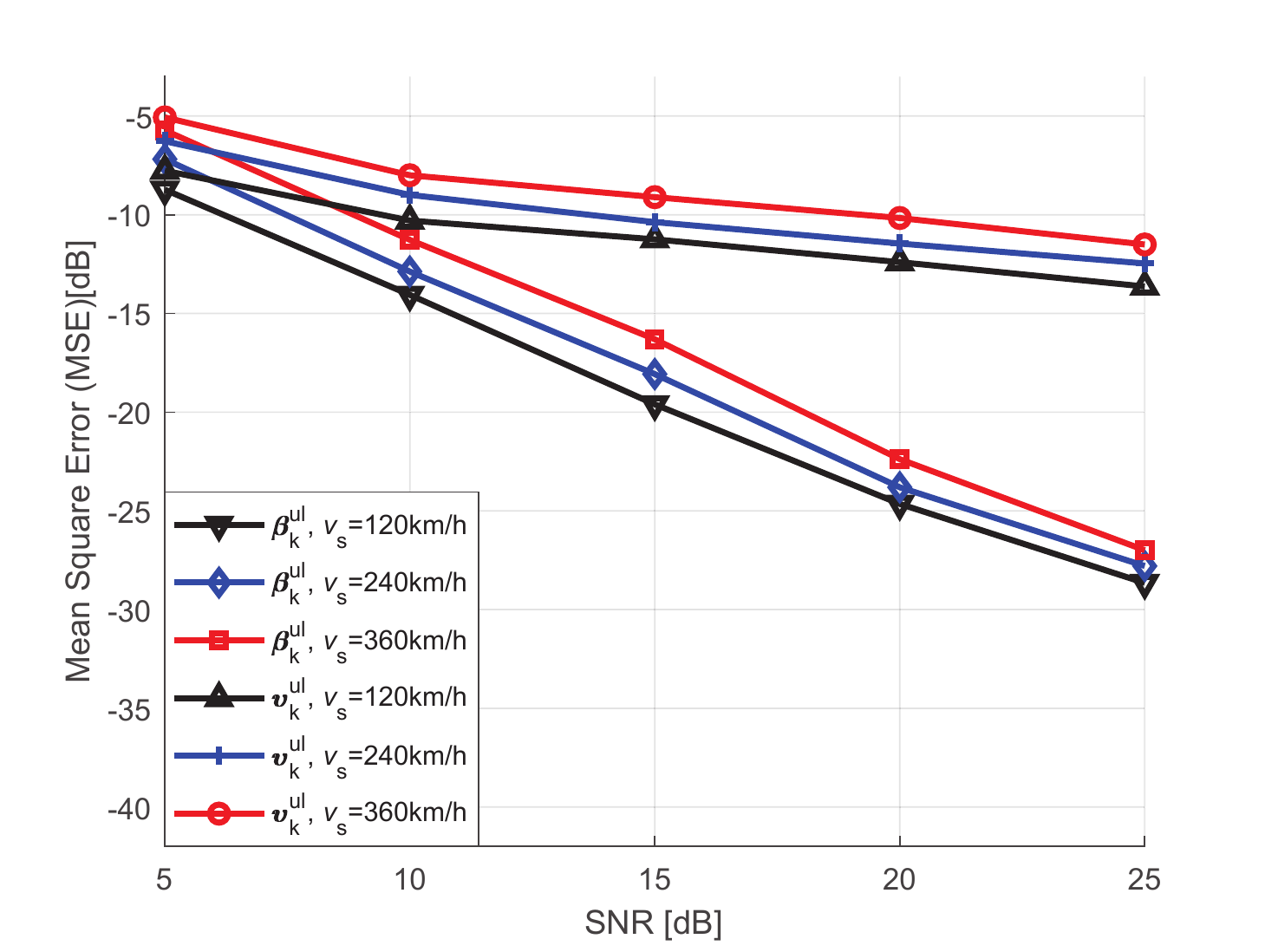}
	\caption{The MSEs  of $\boldsymbol\beta_k^{ul}\!,\!\boldsymbol\upsilon_k^{ul}$ versus SNRs at different users' moving speeds.}
	\label{fig:new_pl_sweep_SNR_change_V_algorithm(1)}
\end{figure}

Fig. \ref{fig:ce_sweep_SNR_change_V} and Fig. \ref{fig:new_pl_sweep_SNR_change_V_algorithm(1)} shows the MSE performance for $\mathbf g_k^{ul}$
and $\boldsymbol{\beta}_k^{ul}$, $\boldsymbol{\upsilon}_k^{ul}$ at different SNRs, where three mobility conditions, i.e., $v_s=120,240,360$ Km/h, are considered. Moreover, we
set the number of paths and observation points as $P=12$ and $N_t=40$, respectively.
Notice that we use the notation ``fast EM-VB" to denote the low complex EM-VB algorithm.
As shown in Fig. \ref{fig:ce_sweep_SNR_change_V} and Fig. \ref{fig:new_pl_sweep_SNR_change_V_algorithm(1)}, the MSE decreases as the SNR increases.
Besides, the bigger the speed $v_s$ is, the higher the MSE curves of $\mathbf g_k^{ul}$, $\boldsymbol{\beta}_k^{ul}$, and $\boldsymbol{\upsilon}_k^{ul}$ are. However,
the difference is very little, which is not the same with the case in \cite{Ma_J_SBL_Time_varing}. This is because that
we directly recover the physical intrinsic parameters $\{\tau_{k,l}^{ul},\upsilon_{k,l}^{ul},\theta_{k,l}^{ul}\}$ but
not the mobility channels.
Furthermore, it can be seen from Fig. \ref{fig:ce_sweep_SNR_change_V} that the performance of our proposed fast EM-VB is only $2\sim3$ dB lower than that of EM-VB, which shows the effectiveness and robustness of fast EM-VB.
Notice that, in Fig. \ref{fig:new_pl_sweep_SNR_change_V_algorithm(1)}, we only present the fast EM-VB performance for the clarity.
In the following, all our present performance curves are
from the fast EM-VB.

\begin{figure}[!htp]
	\centering
	\includegraphics[width=3in]{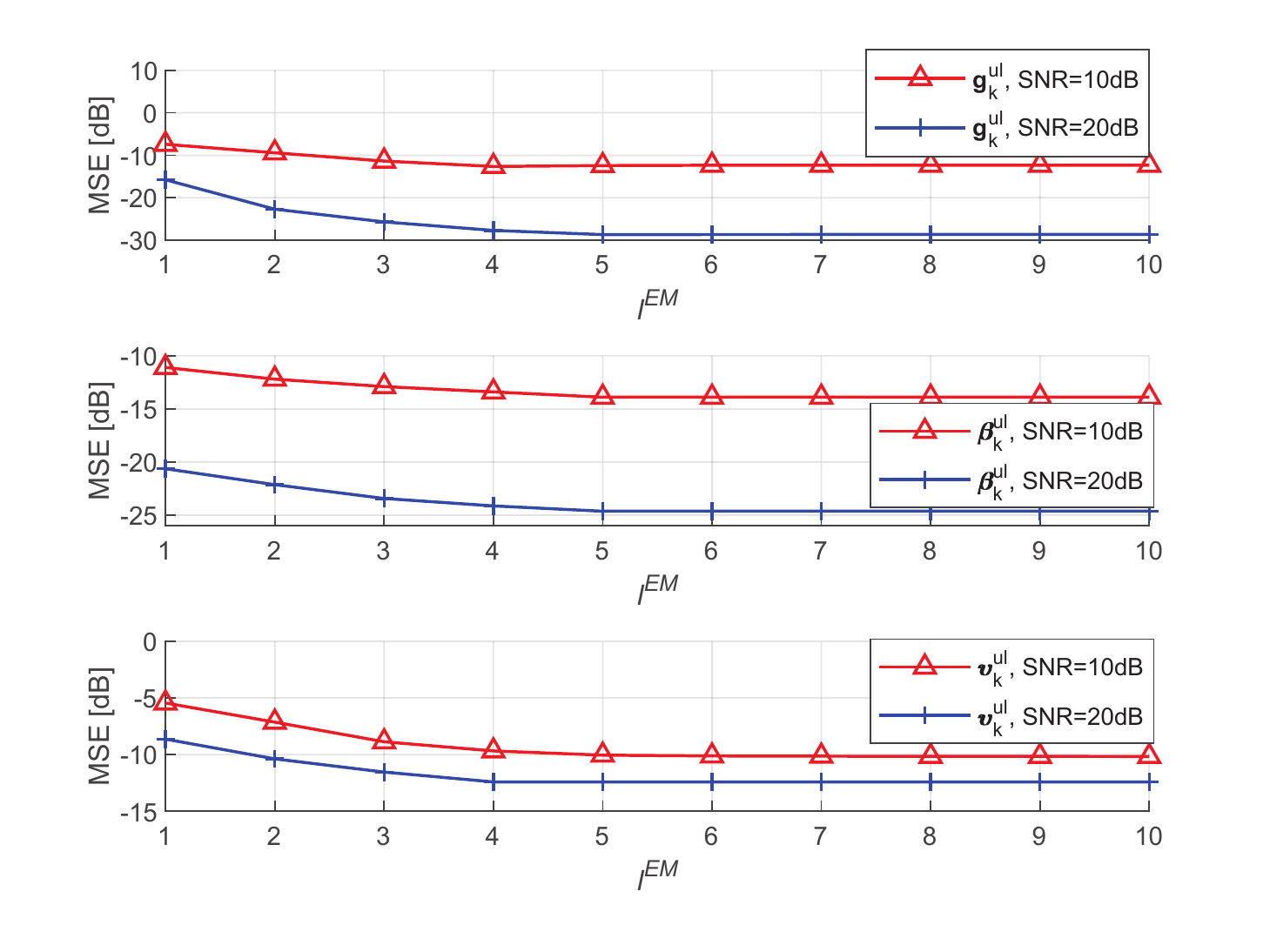}
	\caption{The MSE curves of the channel parameters versus $l^{EM}$.}
	\label{fig:ce_pl_sweep_iter_EM_change_SNR_10_20_K_8_N_t_20}
\end{figure}

In Fig. \ref{fig:ce_pl_sweep_iter_EM_change_SNR_10_20_K_8_N_t_20},
we give the MSE curves of the channel parameters versus the EM iteration index $l^{EM}$.
Two different SNRs, i.e., 10dB and 20dB, are examined. Same with Fig. \ref{fig:ce_sweep_SNR_change_V}, we set
$P=12$, and $N_t=40$. It can be checked from Fig.\ref{fig:ce_pl_sweep_iter_EM_change_SNR_10_20_K_8_N_t_20} only five iterations
are needed for $\boldsymbol{\beta}_k^{ul},\boldsymbol{\upsilon}_k^{ul}$, and $\mathbf g_{k}^{ul}$
to achieve their steady states.


\begin{figure}[!htp]
	\centering
	\includegraphics[width=3in]{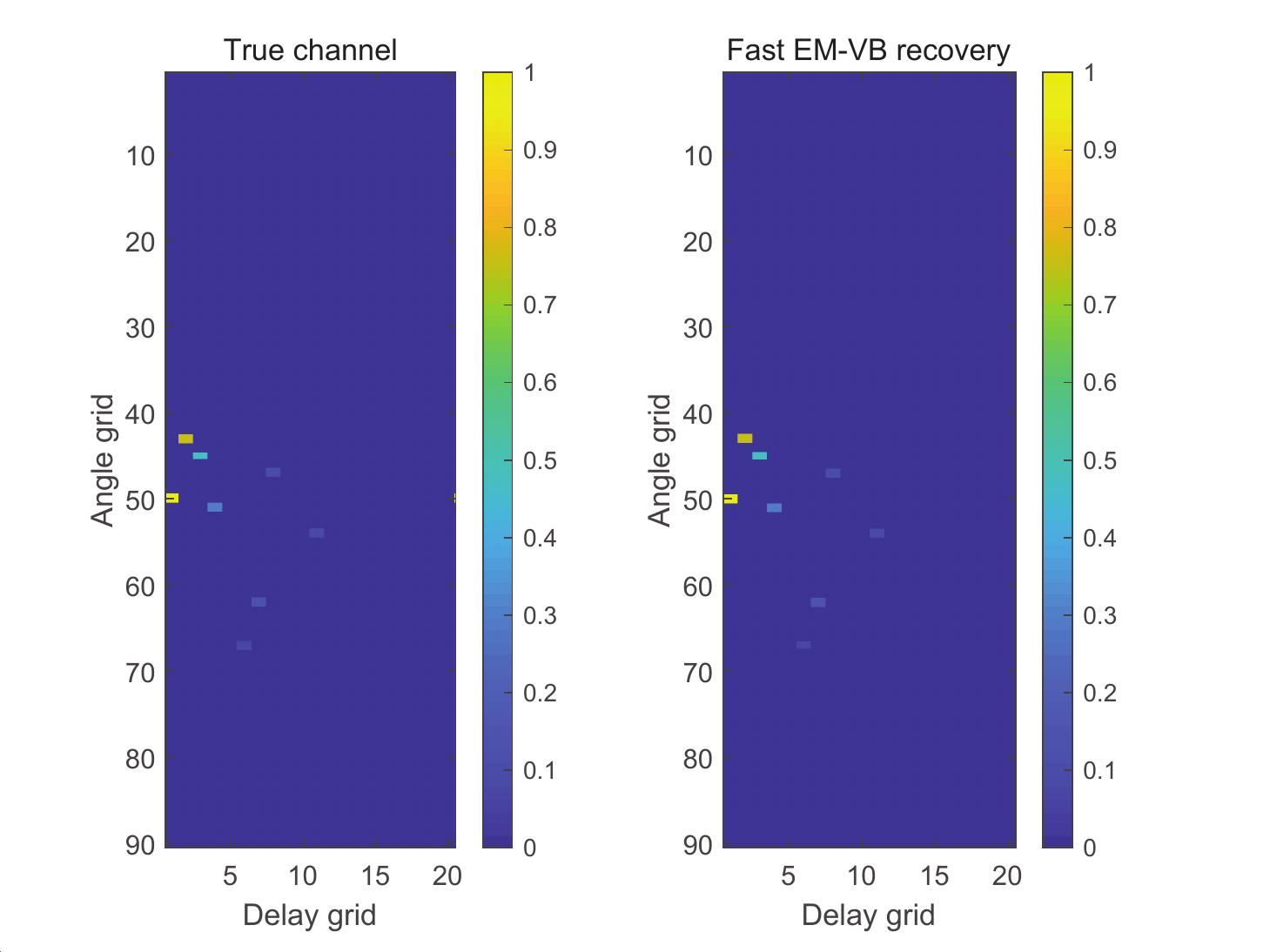}
	\caption{The true and recovered locations of the non-zero elements for $\tilde{\mathbf{G}}_k^{ul}$.}
	\label{fig:channel_support_recovery}
\end{figure}
Fig. \ref{fig:channel_support_recovery} displays the true and estimated locations of the nonzero elements within $\tilde{\mathbf{G}}_k^{ul}$. It can be found that there are only $P=12$ dominant paths for user $k$. Moreover, from the two sub-figures, we can find the accurate support recovery capability of our algorithm.

\begin{figure}[!htp]
	\centering
	\includegraphics[width=3in]{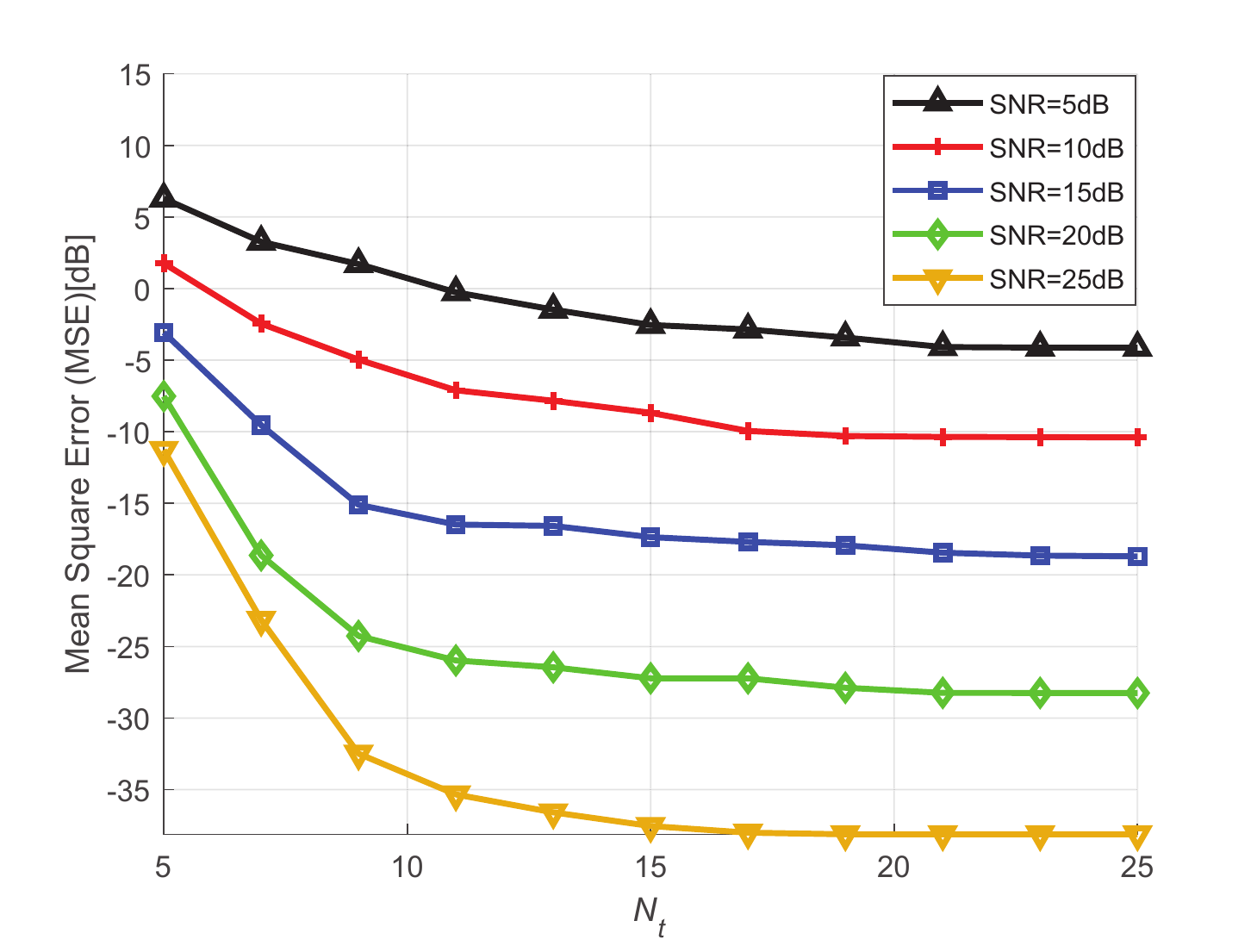}
	\caption{The MSE performance of $\mathbf g^{ul}_k$'s estimation with respect to $N_t$.}
	\label{fig:channel_estimate_sweep_N_t_change_SNR_fix_K_6_step_2}
\end{figure}

In Fig. \ref{fig:channel_estimate_sweep_N_t_change_SNR_fix_K_6_step_2}, we study the MSE performance of $\mathbf g^{ul}_k$'s estimation with respect to the number of observations $N_t$,
where $P=12$, and $\text{SNR}=5,10,15,20,25$ dB.
It can be found that with the number of observations increasing, the MSEs for all SNRs gradually decrease and
would converge to different steady states.
Moreover, the higher the SNR is, the faster the MSE curves would converge. The above observations are not unexpected
and can be explained as follows. Much more observations can help us determine the nonzero locations of $\mathbf g^{ul}_k$
in one easier way, which could enhance the estimation performance of $\mathbf g^{ul}_k$.
However, the MSEs of $\mathbf g^{ul}_k$ are determined by the training power and the noise variance.
With different $P$ at fixed SNR, the training power is fixed. Hence, increasing of $P$ can make sure that
MSEs of  $\mathbf g^{ul}_k$ always decrease.

\begin{figure}[!htp]
	\centering
	\includegraphics[width=3in]{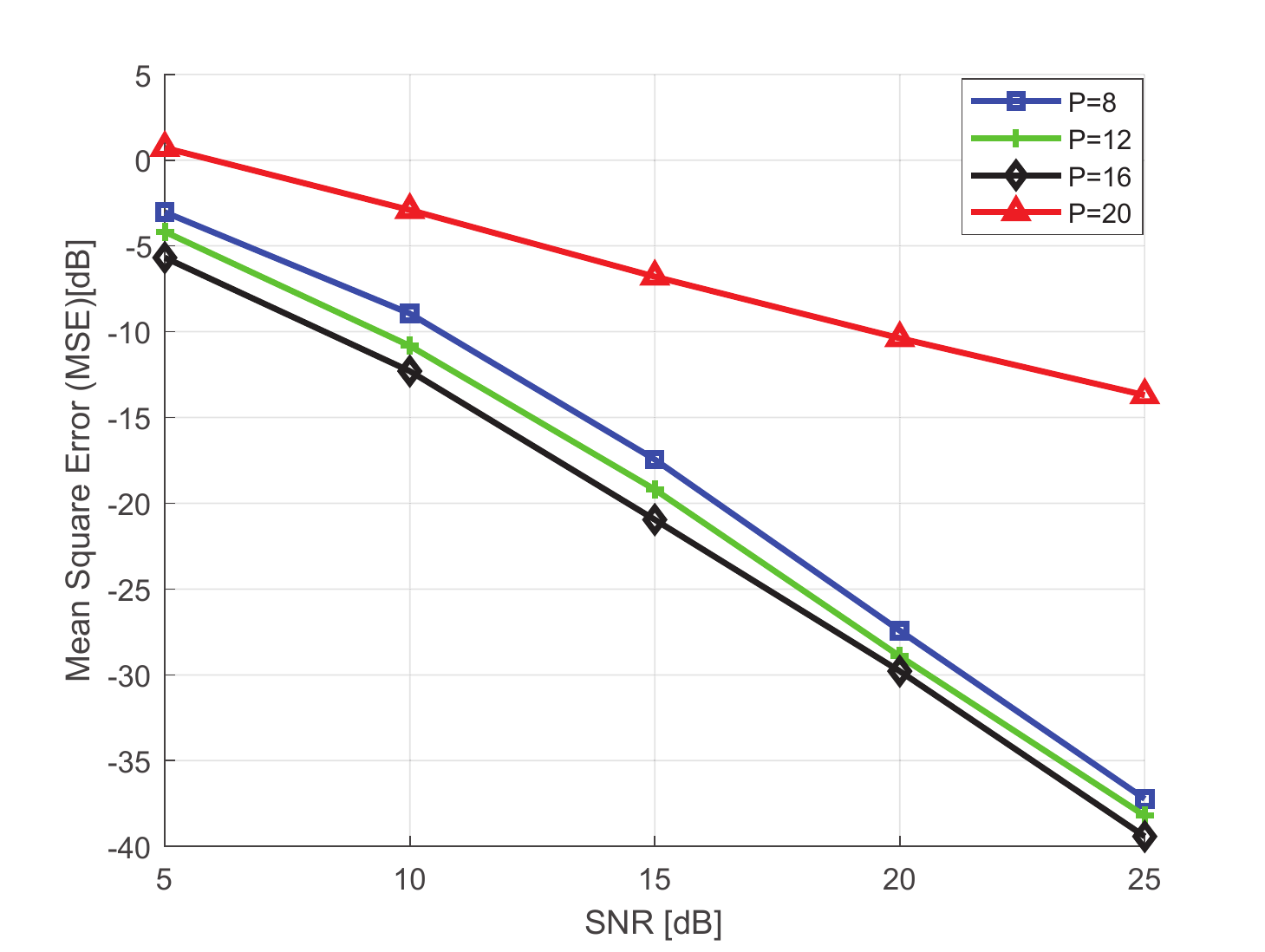}
	\caption{The MSEs of $\mathbf g_k^{ul}$ under different sparsity conditions.}
	\label{fig:channel_estimate_sweep_K_change_SNR_fix_N_t_20}
\end{figure}

In order to examine  the impact of $\mathbf g_k^{ul}$'s sparsity on its estimation performance,
we present $\mathbf g_k^{ul}$'s MSE curves with different $P$ in Fig. \ref{fig:channel_estimate_sweep_K_change_SNR_fix_N_t_20}.
Moreover, $N_t$ is fixed as 18, which means that the maximum unknown parameters that low complex VB-EM can effectively estimate is $18$.
Hence, the bigger $P$ is, the less sparse $\mathbf g_k^{ul}$  will be. From Fig. \ref{fig:channel_estimate_sweep_K_change_SNR_fix_N_t_20}, we have the following observations: If $P$ is smaller than $N_t$,
the MSE curves of  $\mathbf g_k^{ul}$ decrease with the increasing of $P$, which is because
that it is  difficult for the low complex EM-VB to determine the few nonzero locations of $\mathbf g_k^{ul}$.
However, when $P$ becomes bigger than $N_t$, we can not construct enough independent observation equations for unknown nonzero elements in $\mathbf g_k^{ul}$, even if the exact nonzero locations of $\mathbf g_k^{ul}$ are  known.

\begin{figure}[!htp]
	\centering
	\includegraphics[width=3in]{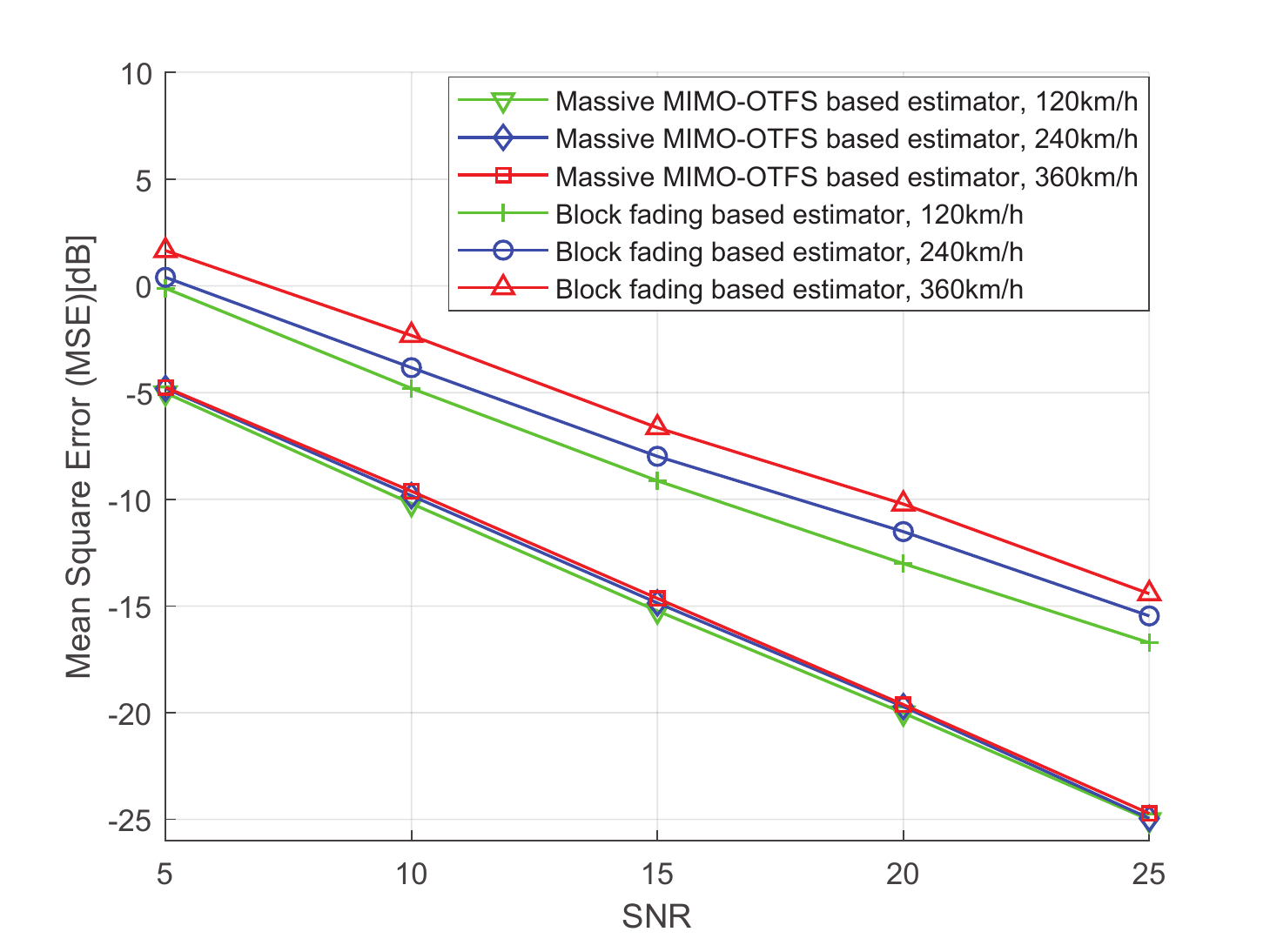}
	\caption{The MSEs of $\bar{\mathbf h}^{no}_k$ under different sparsity conditions.}
	\label{fig:dl_channel_estimate}
\end{figure}

Finally, {we show the MSE curves for the DL delay-Doppler-angle domain $\bar h_{k,i,j,q}$.
Here, the case that different paths of the user $k$ are orthogonal over angle domain is considered.
The FDD model is considered here, where the carrier frequency along the DL is set as $5.98$ GHz.}
Moreover, the corresponding performance for the methods in
\cite{Ma_J_SBL_Time_varing} are also presented for comparison. In order to apply the scheme in
\cite{Ma_J_SBL_Time_varing}, we adopt the CE-BEM to describe the time-varying massive MIMO channels with
BEM coefficient vectors over the time domain, and utilize the framework in  \cite{Ma_J_SBL_Time_varing}  to track
different BEM coefficient vectors in different OFDM symbols. Then, we construct the high mobility channel in the time domain, and use \eqref{eq:fiveB H^DDS}, \eqref{eq:fiveB H^DDA} to obtain the corresponding delay-Doppler-angle domain channel.
From  Fig. \ref{fig:dl_channel_estimate}, we can see that the performance of the massive MIMO-OTFS based DL channel estimator is much better than that of block fading based method. Moreover, similar to Fig. \ref{fig:ce_sweep_SNR_change_V} and Fig. \ref{fig:dl_channel_estimate}, different form the block fading based method, the performance of the massive MIMO-OTFS based DL channel estimator is not closely coupled with the users' mobility speeds.

\section{Conclusion}

In this paper, we examined the  UL-aided high mobility DL channel estimation scheme for the massive MIMO-OTFS networks.
The EM-VB framework was utilized to recover
the UL channel parameters including the angles, the delays, the Doppler frequencies, and the channel gains.
Then, we resorted to the fast Bayesian inference to design one low complex EM-VB. Correspondingly,  the angle, the delay, and the Doppler reciprocity  between UL and DL was fully exploited to
reconstruct
 the parameters
 for DL channels at BS. Furthermore, DL massive MIMO channel estimation over
the delay-Doppler-angle domain was carefully studied.
{Within this phase,
{we
analyzed the channel dispersion of the OTFS over the delay-Doppler domain and
designed three DL 3D channel training schemes according to the scattering characteristics over the 3D angle-delay-Doppler domain.}
Simulation results showed that our proposed strategy is valid and has strong robustness.

\balance

\end{document}